\documentclass[aps,prb,twocolumn,10pt,superscriptaddress,nofootinbib]{revtex4-2}

\usepackage{amsmath, amssymb}
\usepackage{mathtools}
\usepackage{physics}
\usepackage{graphicx}
\usepackage{hyperref}
\usepackage{slashed}
\usepackage{orcidlink}
\usepackage{bbm}
\usepackage{mathtools}
\usepackage{newtxtext}
\usepackage[varvw]{newtxmath}
\usepackage{placeins} % for \FloatBarrier command

\hypersetup{
    colorlinks,
    linkcolor={blue},
    citecolor={blue},
    urlcolor={blue}
}

\usepackage{color}
\usepackage[normalem]{ulem}
\usepackage[export]{adjustbox}
\usepackage{bbold}
\allowdisplaybreaks[1]

\newcommand{\rme}{\mathrm{e}}
\newcommand{\rmi}{\mathrm{i}}
\newcommand{\rmd}{\mathrm{d}}
\newcommand{\Nf}{N_\text{f}}
\newcommand{\Nfc}{N_\text{f,c}}

\begin{document}

%\title{Gross-Neveu-XY quantum criticality in twisted Dirac materials}
\title{Gross-Neveu-XY quantum criticality in moiré Dirac materials}

\author{Bilal Hawashin}
\email{hawashin@tp3.rub.de}
\affiliation{Theoretische Physik III, Ruhr-Universit\"at Bochum, D-44801 Bochum, Germany}

\author{Michael M. Scherer}
\email{scherer@tp3.rub.de}
\affiliation{Theoretische Physik III, Ruhr-Universit\"at Bochum, D-44801 Bochum, Germany}

\author{Lukas Janssen}
\email{lukas.janssen@tu-dresden.de}
\affiliation{Institut für Theoretische Physik and Würzburg-Dresden
Cluster of Excellence ct.qmat, TU Dresden, Dresden, Germany}

\begin{abstract}
Two-dimensional van-der-Waals materials offer a highly tunable platform for engineering electronic band structures and interactions. By employing techniques such as twisting, gating, or applying pressure, these systems enable precise control over the electronic excitation spectrum. In moir\'e bilayer graphene, the tunability facilitates the transition from a symmetric Dirac semimetal phase through a quantum critical point into an interaction-induced long-range ordered phase with a finite band gap. At charge neutrality, the ordered state proposed to emerge from twist-angle tuning is the Kramers intervalley-coherent insulator. In this case, the transition falls into the quantum universality class of the relativistic Gross-Neveu-XY model in 2+1 dimensions. Here, we refine estimates for the critical exponents characterizing this universality class using an expansion around the lower critical space-time dimension of two. We compute the order-parameter anomalous dimension $\eta_\varphi$ and the correlation-length exponent $\nu$ at one-loop order, and the fermion anomalous dimension $\eta_\psi$ at two-loop order. Combining these results with previous findings from the expansion around the upper critical dimension, we obtain improved estimates for the universal exponents in $2+1$ dimensions via Padé interpolation. For $\Nf = 4$ four-component Dirac fermions, relevant to moir\'e bilayer graphene, we estimate $1/\nu=0.916(5)$, $\eta_\varphi=0.926(13)$, and $\eta_\psi=0.0404(13)$. For $\Nf = 2$, potentially relevant to recent tetralayer WSe$_2$ experiments, the Gross-Neveu-XY fixed point may be unstable due to a fixed-point collision at $\Nfc$, with $\Nfc = 1 + \sqrt{2} + \mathcal{O}(\epsilon)$ in the expansion around the lower critical dimension.
\end{abstract}

\maketitle

%%%%%%%%%%%%%%%%%%%%%%%%%%%%%%%%%%%%%%%%%%%%%%%%%%%%%%%%%%%%%%%%%%%%%%%%%%%%%%%%
\section{Introduction} 
%%%%%%%%%%%%%%%%%%%%%%%%%%%%%%%%%%%%%%%%%%%%%%%%%%%%%%%%%%%%%%%%%%%%%%%%%%%%%%%%

Low-energy excitations of various lattice models of hopping electrons have been shown to be effectively described by relativistic fermion field theories. The primary example is given by the tight-binding model on the honeycomb lattice~\cite{Katsnelson:2007}, in which case the low-energy quasiparticles behave as relativistic Dirac fermions in $2+1$ dimensions. The vanishing density of states in combination with the lattice symmetries protects the Dirac quasiparticles from acquiring a spectral gap, rendering the Dirac semimetal a stable gapless phase at weak interactions. However, if interactions exceed a certain finite threshold, the system can undergo one of several possible instabilities, leading to long-range order. These include antiferromagnetic order~\cite{Sorella:1992,Martelo:1996,Paiva:2005,HerbutGraphene:2006,assaad13}, charge-density wave order~\cite{raghu08,HerbutInteracting:2009,weeks10}, Kekul\'e valence-bond solid order~\cite{Scherer:2016,LiFermi:2017,Xu:2018}, or superconducting order~\cite{RoyKekule:2010,Otsuka:2018}. The presence of order is described by an order parameter and is field-theoretically implemented as a bosonic field that is coupled to the Dirac fermions via a Yukawa interaction. Models of interacting Dirac fermions typically fall into universality classes of the Gross-Neveu type~\cite{HerbutInteracting:2009,Gehring:2015,Zerf4L:2017,herbut24}.

In recent years, a highly tunable platform for realizing various strongly-correlated phases has emerged, involving the stacking and twisting of two layers of graphene, known as twisted bilayer graphene~\cite{Shallcross:2010,Bistritzer:2011,Luican:2011,Sanchez:2012,Cao:2016}. Upon tuning the twist angle to the ``magic angle'' of approximately $1.1^\circ$~\cite{Bistritzer:2011,Lopes:2012,Tarnopolsky:2019}, the bands become nearly flat and exhibit a topological character~\cite{Bernevig:2021}. Experimentally, a complex phase diagram has emerged, featuring a variety of correlated phases, including correlated insulators and unconventional superconductors~\cite{Lu:2019,Cao:2018}. At charge neutrality, the insulating ground state is theoretically predicted to feature Kramers intervalley-coherent order~\cite{Po:2018,Bultinck:2020,kwan21,hofmann22,rai24}. Increasing the twist angle away from the magic angle is expected to destabilize the Kramers intervalley-coherent order, resulting in a reduction of the interaction-induced band gap, culminating in a quantum critical point at around $1.2^\circ$~\cite{Biedermann:2024,Huang:2024}. For larger twist angles beyond this critical value, the Dirac semimetal state is believed to be stable. This semimetal-to-insulator transition breaks the U(1)$_\text{valley}$ symmetry of system, and can thus be captured by an XY order parameter. Due to the presence of gapless fermions, the quantum phase transition goes beyond the Landau-Ginzburg-Wilson paradigm, which relies solely on an order-parameter description~\cite{vojta18}. Instead, it has been suggested that the transition belongs to the $(2+1)$-dimensional Gross-Neveu-XY universality class~\cite{Biedermann:2024,Huang:2024}.

Field theories that fall into Gross-Neveu-type universality classes can be defined within purely fermionic as well as fermion-boson formulations. The different formulations are related by a Hubbard-Stratonovich transformation~\cite{Herbut:2007}. The purely fermionic formulations are referred to as Gross-Neveu models, while the fermion-boson formulations are typically referred to as Gross-Neveu-Yukawa models. On the technical side, the Yukawa coupling in Gross-Neveu-Yukawa models becomes marginal precisely in $D=3+1$ space-time dimensions, which is identified as upper critical dimension. Similarly, four-fermion interactions in Gross-Neveu models become marginal in $D=1+1$ space-time dimensions, which defines the lower critical dimension. Slightly below (above) to the upper (lower) critical dimension $D_\mathrm{c}$, the theory becomes perturbatively renormalizable and admits a systematic expansion in powers of $|D-D_\mathrm{c}|$. This approach has been successfully applied to the Gross-Neveu-XY universality class in an expansion around the upper critical dimension, up to four-loop order~\cite{Zerf4L:2017}. Estimates for critical exponents have also been accessed with other methods, such as the large-$\Nf$ expansion~\cite{GraceyN3XY:2021}, the functional renormalization group~\cite{ClassenXY:2017,JanssenXYFRG:2018,Tolosa:2025} or quantum Monte Carlo simulations~\cite{LiFermi:2017,Xu:2018,Otsuka:2018,Huang:2024}. 

A significant improvement of estimates from perturbative expansions around the upper critical dimension can be obtained by a suitable interpolation with results from the lower critical dimension, as has been shown for the Gross-Neveu-Ising~\cite{JanssenAFM:2014, IhrigIsing:2018} as well as the Gross-Neveu-Heisenberg case~\cite{Ladovrechis:2023}. 
However, for the Gross-Neveu-XY universality class, the required results from the expansion around the lower critical dimension are still absent in the literature.
A technical difficulty that arises in renormalization group~(RG) calculations near two space-time dimensions is that all four-fermion terms compatible with the symmetries of the model are potentially important and can, in general, be generated under the RG. An unbiased study therefore necessitates the inclusion of all symmetry-allowed and RG relevant and marginal operators in the action. While the Gross-Neveu-Ising four-fermion interaction is closed under the RG~\cite{Gehring:2015, gracey16}, the Gross-Neveu-Heisenberg model, for example, is not~\cite{Ladovrechis:2023}.

%%%%%%%%%%%%%%%%%%%%%%%%%%%%%%%%%%%%%%%%%%%%%%%%%%%%%%%%%%%%%%%%%%%%%%%%%%%%%%%%
\begin{table}[t!]
\caption{%
Gross-Neveu-XY critical exponents for $\Nf = 4$ four-component Dirac fermions in 2+1 space-time dimensions, relevant for the twist-tuned transition in moir\'e bilayer graphene~\cite{Biedermann:2024, Huang:2024}, from interpolation between series expansions near lower and upper critical dimensions (this work) in comparison with results from fourth-order $4-\varepsilon$ expansion~\cite{Zerf4L:2017}, functional RG in local potential approximation~\cite{ClassenXY:2017}, second-order (third-order for $\eta_\psi$) $1/\Nf$ expansion~\cite{GraceyN3XY:2021}, and quantum Monte Carlo (QMC) simulations of a momentum-space-discretized continuum model~\cite{Huang:2024}.}
\begin{tabular*}{\linewidth}{@{\extracolsep{\fill}} l l l l l}
\noalign{\smallskip}\hline\hline\noalign{\smallskip}
 $\Nf = 4$ & \multicolumn{1}{c}{Year} & \multicolumn{1}{c}{$1/\nu$} & \multicolumn{1}{c}{$\eta_\varphi$} & \multicolumn{1}{c}{$\eta_\psi$}\\
\noalign{\smallskip}\hline\noalign{\smallskip}
 Interpolation (this work) & {2025} & {0.916(5)} & {0.926(13)} & {0.0404(13)} \\
 $4-\varepsilon$ expansion~\cite{Zerf4L:2017} & 2017 & 0.94(8) & 0.99(5) & 0.043(5) \\
 Functional RG~\cite{ClassenXY:2017} & 2017 & 0.92 & 0.95 & 0.027 \\
 Large-$\Nf$ expansion~\cite{GraceyN3XY:2021} & 2021 & 0.905(33) & 0.941(4) & 0.0376(6)   \\
 QMC, momentum grid~\cite{Huang:2024} & 2024 & 0.86(4) & 0.72(19) & \multicolumn{1}{c}{--} \\
\noalign{\smallskip} \hline \hline
\end{tabular*}
\label{tab:summaryNf=4}
\end{table}
%%%%%%%%%%%%%%%%%%%%%%%%%%%%%%%%%%%%%%%%%%%%%%%%%%%%%%%%%%%%%%%%%%%%%%%%%%%%%%%%

In this work, we provide an RG study of the Gross-Neveu-XY theory space close to the lower critical dimension. We determine a Fierz-complete basis of four-fermion interactions and calculate the beta functions at one-loop order. We identify a two-dimensional one-loop closed subspace that includes a critical fixed point associated with the Gross-Neveu-XY universality class, and determine the correlation-length exponent $1/\nu$ and the order-parameter anomalous dimension $\eta_\varphi$ to first order in $\epsilon$, and the fermion anomalous dimension $\eta_\psi$ to second order in $\epsilon$.
We find that, close to two space-time dimensions, the Gross-Neveu-XY fixed point is stable only above a critical flavor number of $\Nfc = 1+ \sqrt{2} + \mathcal O(\epsilon)$ four-components fermions. 
For $\Nf < \Nfc$ and without additional fine-tuning, a lattice model realizing a Gross-Neveu-XY quantum phase transition falls into the universality class defined by a different quantum critical fixed point, with exponents differing from those of the (now unstable) Gross-Neveu-XY fixed point.
Since $\Nf = 4$, relevant for the twist-tuned transition in moiré bilayer graphene~\cite{Biedermann:2024, Huang:2024}, lies well above $\Nfc$ in the expansion near the lower critical dimension, we expect the Gross-Neveu-XY fixed point to govern the Gross-Neveu-XY transition in the physical space-time dimension $D=2+1$.
For $\Nf = 2$, which is relevant for the semimetal-to-superconductor transition on the triangular lattice~\cite{Otsuka:2018} and the semimetal-to-Kekulé-insulator transition on the honeycomb lattice~\cite{LiFermi:2017, Xu:2018}, higher-order terms are needed to reliably determine whether the Gross-Neveu-XY fixed point or an alternative fixed point, which we identify in this work, governs the transition in a generic Dirac system that realizes a transition to an ordered state with broken U(1) symmetry.
This scenario may also be relevant for recent experiments on tetralayer WSe$_2$, which observed a twist-tuned semimetal-to-insulator transition~\cite{ma24}. However, the order parameter has not yet been unambiguously identified in this case.

For the Gross-Neveu-XY fixed point, we obtain improved estimates by using a Padé approximant interpolation scheme, bridging our results with previous four-loop calculations from the expansion around the upper critical dimension~\cite{Zerf4L:2017} for various flavor numbers relevant to different lattice realizations. We present our numerical results alongside comparisons with previous estimates in Tables~\ref{tab:summaryNf=4} and \ref{tab:summaryNf=2}.

The remainder of this paper is organized as follows: In Sec.~\ref{sec:chiralXY}, we introduce the Gross-Neveu-XY model. The RG analysis in $2+\epsilon$ space-time dimensions is presented in Sec.~\ref{sec:rg}. In Sec.~\ref{sec:critexp}, we discuss our results for the critical exponents governing the quantum critical behavior of twisted Dirac materials. Our conclusions are provided in  Sec.~\ref{sec:conclusions}. 
A discussion of an alternative formulation of the Gross-Neveu-XY model and additional data for critical exponents are provided in two appendices.

%%%%%%%%%%%%%%%%%%%%%%%%%%%%%%%%%%%%%%%%%%%%%%%%%%%%%%%%%%%%%%%%%%%%%%%%%%%%%%%%
\begin{table}[t!]
\caption{%
Same as Table~\ref{tab:summaryNf=4}, but for $\Nf = 2$ four-component Dirac fermions, relevant for the semimetal-to-superconductor transition on the triangular lattice~\cite{Otsuka:2018} and the semimetal-to-Kekulé-insulator transition on the honeycomb lattice~\cite{LiFermi:2017, Xu:2018}.
Note that the results from interpolation (this work), $4-\varepsilon$ expanion~\cite{Zerf4L:2017}, large-$\Nf$ expansion~\cite{GraceyN3XY:2021}, and functional RG~\cite{ClassenXY:2017} come with an important caveat:
If $\Nfc > 2$ in $D = 2+1$, as is the case near the lower critical dimension, the presented exponents are accessible in a lattice model only with additional fine-tuning. A generic Gross-Neveu-XY quantum phase transition, such as those studied in Refs.~\cite{LiFermi:2017, Xu:2018, Otsuka:2018}, would instead fall into the universality class defined by the quantum critical fixed point in Eq.~\eqref{eq:GNXY-FP-Nf=2}, with exponents differing from those presented in the first four rows of this table.
}
\begin{tabular*}{\linewidth}{@{\extracolsep{\fill}} l l l l l}
\noalign{\smallskip}\hline\hline\noalign{\smallskip}
 $\Nf = 2$ & \multicolumn{1}{c}{Year} & \multicolumn{1}{c}{$1/\nu$} & \multicolumn{1}{c}{$\eta_\varphi$} & \multicolumn{1}{c}{$\eta_\psi$}\\
\noalign{\smallskip}\hline\noalign{\smallskip}
 Interpolation (this work) & {2025} & {0.904(9)} & {0.850(27)} & {0.095(19)}\\
 $4-\varepsilon$ expansion~\cite{Zerf4L:2017} & 2017 & 0.94(15) & 0.94(19) & 0.118(8) \\
 Large-$\Nf$ expansion~\cite{GraceyN3XY:2021} & 2021 & 0.84(8) & 0.90(2) & 0.082(6) \\
 Functional RG~\cite{ClassenXY:2017} & 2017 & 0.86 & 0.88 & 0.062 \\
 QMC, honeycomb~\cite{LiFermi:2017} & 2017 & 0.94(4) & 0.71(3) & \multicolumn{1}{c}{--} \\
 QMC, honeycomb~\cite{Xu:2018} & 2018 & 0.95(5) & 0.75(13) & \multicolumn{1}{c}{--} \\
 QMC, triangular~\cite{Otsuka:2018} & 2018 & 0.932(6) & 0.64(2) & 0.151(4) \\
\noalign{\smallskip}\hline\hline
\end{tabular*}
\label{tab:summaryNf=2}
\end{table}
%%%%%%%%%%%%%%%%%%%%%%%%%%%%%%%%%%%%%%%%%%%%%%%%%%%%%%%%%%%%%%%%%%%%%%%%%%%%%%%%

%%%%%%%%%%%%%%%%%%%%%%%%%%%%%%%%%%%%%%%%%%%%%%%%%%%%%%%%%%%%%%%%%%%%%%%%%%%%%%%%
\section{Gross-Neveu-XY model} 
\label{sec:chiralXY}
%%%%%%%%%%%%%%%%%%%%%%%%%%%%%%%%%%%%%%%%%%%%%%%%%%%%%%%%%%%%%%%%%%%%%%%%%%%%%%%%

We consider the Gross-Neveu-XY model in $D$ Euclidean space-time dimensions which is described by the microscopic action $S = \int \rmd^Dx\, \mathcal{L}$ and
\begin{align} \label{eq:GNXYaction}
    \mathcal{L} = \bar{\psi}^\alpha \gamma_\mu \partial_\mu \psi^\alpha + \frac{g}{2 \Nf} \big[(\bar{\psi}^\alpha \gamma_3 \psi^\alpha)^2 + (\bar{\psi}^\alpha \gamma_5 \psi^\alpha)^2 \big].
\end{align}
Here, $\mu = 0,\dots,D-1$ denotes the space-time index, $\alpha = 1,\dots,\Nf$ the flavor index, and $\bar{\psi} = \psi^\dagger \gamma_0$ represent the Dirac conjugate of the spinor $\psi$. The summation convention over repeated indices is assumed. We choose a four-dimensional representation of the Clifford algebra, $\{\gamma_\mu, \gamma_\nu\} = 2 \delta_{\mu \nu} \mathbb{1}_4$, and hence each flavor $\psi^\alpha$ has four spinor components. 
Below $3+1$ space-time dimensions, this representation is reducible.
In $D=1+1$ space-time dimensions, this results in three additional Hermitian matrices that anticommute with both $\gamma_\mu$ and square to one, namely, $\gamma_2$, $\gamma_3$, and $\gamma_5$.
In $D=2+1$ space-time dimensions, $\gamma_3$ and $\gamma_5$ remain.
An explicit representation of the gamma matrices is given by $\gamma_0 = \mathbb{1}_2 \otimes \sigma_z, \gamma_1 = \sigma_z \otimes \sigma_y, \gamma_2 = \mathbb{1}_2 \otimes \sigma_x, \gamma_3 = \sigma_x \otimes \sigma_y$ and $\gamma_5 = \sigma_y \otimes \sigma_y$, with the Pauli matrices $\sigma_{x,y,z}$.
The theory defined by the Lagrangian in Eq.~\eqref{eq:GNXYaction} features  a quantum critical point in $2 < D < 4$ space-time dimensions that falls into the Gross-Neveu-XY universality class, which has previously been accessed within a $4 - \varepsilon$ expansion~\cite{Zerf4L:2017}, a large-$\Nf$ expansion~\cite{GraceyN3XY:2021}, functional RG~\cite{ClassenXY:2017,JanssenXYFRG:2018,Tolosa:2025}, and quantum Monte Carlo simulations~\cite{LiFermi:2017, Otsuka:2018, Xu:2018, Huang:2024}. For $\Nf = 2$, the above theory describes the continuous transition from a Dirac semimetal to Kekulé valence bond solid order on the honeycomb lattice~\cite{HerbutGraphene:2006,HerbutInteracting:2009,RoyKekule:2010}. For $\Nf=4$, it describes the twist-tuned transition towards Kramers intervalley-coherent order in moir\'e bilayer graphene~\cite{Huang:2024,Biedermann:2024}. 

In the following, we present an RG analysis based on an expansion around the lower critical dimension, which is complementary to the expansion around the upper critical dimension.
To this end, it is important to discuss the symmetries of the model  at the lower critical space-time dimension $D_\mathrm{c}^\mathrm{low} = 1+1$. 
The set of all fermion models that are invariant under these symmetries defines the Gross-Neveu-XY theory space.

\paragraph{Relativistic symmetry:} Under Lorentz transformations in two Euclidean space-time dimensions, the space-time coordinate transforms as $x_\mu \mapsto x'_\mu = (\Lambda^{-1})_{\mu \nu} x_\nu$, with rotation matrix $\Lambda \in \mathrm{O}(2)$. For the Dirac spinors, this implies
\begin{align}
    \psi^\alpha(x) \mapsto \mathcal S(\Lambda) \psi^\alpha(x'),
\end{align}
where
$\mathcal S(\Lambda) = \rme^{{\omega} \gamma_{0} \gamma_{1}/2}$ and $\omega$ is chosen such that $\mathcal S^{-1}(\Lambda) \gamma_\mu \mathcal S(\Lambda) = \Lambda_{\mu\nu} \gamma_\nu$.

\paragraph{SU(\(\Nf\)) flavor symmetry:} The flavor symmetry acts on the Dirac spinors as
\begin{align}
    \psi^\alpha \mapsto U^{\alpha \beta} \psi^\beta,
\end{align}
with the unitary matrix $U \in \mathrm{SU}(\Nf)$, which is generated by the generalized $\Nf \times \Nf$ Gell-Mann matrices $\{\lambda_1,\dots,\lambda_{\Nf^2 - 1}\}$. 

\paragraph{U(1) charge conservation:} The U(1) symmetry associated with charge conservation reads
\begin{align}
    \psi^\alpha \mapsto \rme^{\rmi \varphi} \psi^\alpha,
\end{align}
with possibly flavor-dependent angle $\varphi \equiv \varphi_\alpha$.

\paragraph{U(1) continuous chiral symmetry:} The use of a reducible representation of the Clifford algebra allows us to define a chirality, using the projector $P_{35}^{\pm} = \frac12 (\mathbb{1} \pm \gamma_{35})$ with $\gamma_{35} = \rmi \gamma_3 \gamma_5$. The Gross-Neveu-XY model is invariant under the associated continuous chiral U(1) symmetry, defined as
\begin{align}
    \psi^\alpha \mapsto \rme^{\rmi \theta \gamma_{35}} \psi^\alpha,
\end{align}
with possibly flavor-dependent angle $\theta \equiv \theta_\alpha$. Under an infinitesimal transformation, ${\delta_\theta \psi^\alpha = \rmi \gamma_{35} \theta \psi^\alpha}$, the real two-tuple $\vec{\varphi} = \rmi (\bar{\psi}^\alpha \gamma_3 \psi^\alpha, \bar{\psi}^\alpha \gamma_5 \psi^\alpha)^\top$ transforms as an O(2) vector,
\begin{align}
\delta_\theta \begin{pmatrix}
        \bar{\psi}^\alpha \gamma_3 \psi^\alpha \\
        \bar{\psi}^\alpha \gamma_5 \psi^\alpha
    \end{pmatrix} = \begin{pmatrix}
        0 & -2\theta \\
        2\theta & 0
    \end{pmatrix} \begin{pmatrix}
        \bar{\psi}^\alpha \gamma_3 \psi^\alpha \\
        \bar{\psi}^\alpha \gamma_5 \psi^\alpha
    \end{pmatrix},
\end{align}
such that the interaction term, which can be written as a scalar product $(\bar\psi^\alpha \gamma_3 \psi^\alpha, \bar\psi^\alpha \gamma_5 \psi^\alpha)^2$, is invariant under the continuous chiral symmetry.

Charge, continuous chiral and flavor symmetry together form a $\mathrm{U}(\Nf) \times \mathrm{U}(\Nf)$ global symmetry with the $2\Nf^2$ generators ${\{\lambda_i\}_{i=1,\dots,\Nf^2} \otimes \{\mathbb{1}_4, \gamma_{35}\}}$ with $\lambda_{\Nf^2} := \mathbb{1}_{\Nf}$.

In addition to the above continuous symmetries, the Gross-Neveu-XY model also features various discrete symmetries, of which we list a few:

\paragraph{$\mathbb{Z}_2$ chiral symmetry:} In addition to the continuous chiral symmetry, the model features a discrete chiral symmetry, which acts on the Dirac spinors as
\begin{align}\label{eq:Z2symm}
    \psi^\alpha &\mapsto \gamma_5 \psi^\alpha, & \bar{\psi}^\alpha & \mapsto -\bar{\psi}^\alpha \gamma_5.
\end{align}    
On the honeycomb lattice, it corresponds to the sublattice-exchange symmetry~\cite{HerbutInteracting:2009, JanssenAFM:2014}.
    
\paragraph{Spatial-inversion symmetry:} Under spatial inversion, the Dirac spinor transforms as  
\begin{align}
    \psi^\alpha(x) \mapsto \mathcal I \psi^\alpha(x'),
\end{align}
with the inversion operator $\mathcal I = \gamma_0$ and $(x'_0,x'_1) = (x_0,-x_1)$.
    
\paragraph{Time-reversal symmetry:} Under Euclidean time-reversal symmetry, we have
\begin{align}\label{eq:TRsymm}
    \psi^\alpha(x) \mapsto \mathcal T \psi^\alpha(x),
\end{align}
with the time-reversal operator $\mathcal T = \rmi \gamma_1 \gamma_5 \mathcal K$, where $\mathcal K$ denotes complex conjugation. Here, we have assumed a representation of the Clifford algebra in which $\gamma_0$ is real.

In Appendix~\ref{app:spinXY}, we present an alternative definition of a Gross-Neveu-XY model, obtained via adding an XXZ spin anisotropy to the SU(2)-symmetric Gross-Neveu-Heisenberg model~\cite{Ladovrechis:2023}. In this formulation, the U(1) continuous chiral symmetry can be understood as residual XY spin rotational symmetry in the presence of the anisotropy.
We explicitly show that there is a one-to-one correspondence between the operators in the theory space of this ``Gross-Neveu-Spin-XY model'' and those of the original Gross-Neveu-XY model defined in Eq.~\eqref{eq:GNXYaction}.
However, we emphasize that this result alone does not yet imply that the two models feature the same critical behavior. This is because the two different models define different trajectories in the same theory space. At criticality, they fall into the domains of attraction of two different Gross-Neveu-XY-type fixed points. We nevertheless demonstrate below that these two critical fixed points share the same universal exponents at the one-loop level. The question of whether this property also holds at higher loop orders represents an interesting direction for future work.

%%%%%%%%%%%%%%%%%%%%%%%%%%%%%%%%%%%%%%%%%%%%%%%%%%%%%%%%%%%%%%%%%%%%%%%%%%%%%%%%
\section{Renormalization group}
\label{sec:rg}
%%%%%%%%%%%%%%%%%%%%%%%%%%%%%%%%%%%%%%%%%%%%%%%%%%%%%%%%%%%%%%%%%%%%%%%%%%%%%%%%

Generically, four-fermion couplings $g$ in $D$ space-time dimensions have mass dimension $[g] = D - 2$. Therefore, exactly in $D=2$, the coupling becomes marginal and the theory perturbatively renormalizable, admitting a systematic expansion in powers of $\epsilon=D-2$. This identifies $D_\mathrm{c}^\mathrm{low}=2$ as the lower critical dimension of relativistic four-fermion theories~\cite{zinnjustin91, juricic09, gracey16, Gehring:2015, Ladovrechis:2023}. For space-time dimensions above $D=2$, the four-fermion coupling $g$ is perturbatively irrelevant and a small $g$ flows towards the noninteracting fixed point in the infrared. A sizable $g$ beyond a certain finite threshold, however, can induce an instability, signalled by a runaway flow towards positive or negative infinite $g$.
During the RG flow, all operators compatible with the symmetries of the model can be generated and in principle need to be taken into account.
For small deviations from the lower critical dimensions, however, interaction terms involving more than four fermion fields are RG irrelevant at any weakly-coupled fixed point.
In order to correctly identify the critical behavior in an unbiased way, it is thus necessary to classify all symmetry-allowed operators up to fourth order in the fermion fields.

\subsection{Symmetry-allowed operators}

The U(1) charge conservation symmetry implies that any allowed term consists of an equal number of spinors~$\psi$ and corresponding Dirac conjugates~$\bar\psi$.
Together with the $\mathrm{SU}(\Nf)$ flavor symmetry,
this leaves us to discuss bilinears of the form $(\bar{\psi}^\alpha \mathcal{O} \psi^\alpha)$
and four-fermion terms of flavor-singlet structure $(\bar{\psi}^\alpha \mathcal{O} \psi^\alpha)(\bar{\psi}^\beta \mathcal{Q} \psi^\beta)$ or flavor non-singlet structure $(\bar{\psi}^\alpha \mathcal{O} \psi^\beta)(\bar{\psi}^\beta \mathcal{Q} \psi^\alpha)$ structure, with $\mathcal{O}$ and $\mathcal{Q}$ are $4 \times 4$ matrices.
However, four-fermion terms with flavor-non-singlet structure can be expressed in terms of terms with flavor-singlet structure through Fierz identities~\cite{HerbutInteracting:2009, JanssenThirring:2010, Gehring:2015}.
A Fierz-complete basis of four-fermion terms can therefore be given by restricting to terms with flavor-singlet structure.
Note that in the special case of $\Nf = 1$, in which case the flavor symmetry is trivial, there exist additional relations between four-fermion terms~\cite{Gehring:2015}, which lead to a smaller number of linearly independent interactions.
In the following, we restrict our discussion to the case $\Nf > 1$, allowing us to focus on terms with a flavor-singlet structure. For simplicity, we omit the explicit notation of the flavor index.

A basis in the 16-dimensional space of $4 \times 4$ operators $\mathcal{O}$ is given by 
\begin{align}
    \mathcal{B} = \{\mathbb{1}_4,\gamma_i, \gamma_{ij}\},
\end{align}
where $\gamma_i$, $i=0,1,2,3,5$ are the five gamma matrices and ${\gamma_{ij} = \frac{\rmi}{2}[\gamma_i,\gamma_j]}$ with $i<j$, $i,j=0,1,2,3,5$, are the corresponding ten independent products of gamma matrices.

Any bilinear $(\bar{\psi} \mathcal{O} \psi)$ with $\mathcal{O} \in \mathcal{B}$ is odd at least under one of the discrete symmetries~\eqref{eq:Z2symm}-\eqref{eq:TRsymm}, and hence no bilinear term, such as a mass term, is allowed by symmetry.
Turning to four-fermion terms $(\bar{\psi} \mathcal{O} \psi)(\bar{\psi} \mathcal{Q} \psi)$, we first note that the relativistic and discrete symmetries listed above imply that $\mathcal O = \mathcal Q$, leaving us with 16 possible four-fermion terms.
These may be further restricted.
First, recall that $\vec{\varphi} = \rmi (\bar{\psi} \gamma_3 \psi, \bar{\psi} \gamma_5 \psi)^\top$ transforms as an O(2) vector under the continuous chiral symmetry.
This implies that the terms with $\mathcal O = \gamma_3$ and $\mathcal O = \gamma_5$ are symmetry-allowed only if they appear in the combination ${(\bar\psi \gamma_3 \psi)^2 + (\bar\psi \gamma_5 \psi)^2}$. An analogous argument holds for terms with $\mathcal O = \mathcal O' \gamma_3$ and $\mathcal O = \mathcal O' \gamma_5$ for operators $\mathcal O'$ that commute with $\gamma_{35}$, such as $\mathcal O' = \gamma_2$.
Similarly, pairs of terms including operators that anticommute with $\gamma_{01}$, such as $\mathcal O_0 = \gamma_0$ and $\mathcal O_1 = \gamma_1$, or $\mathcal O_0 = \gamma_{02}$ and $\mathcal O_1 = \gamma_{12}$, must appear in the combination $(\bar\psi \mathcal O_0 \psi)^2 + (\bar\psi \mathcal O_1 \psi)^2$, to ensure relativistic invariance.
In sum, a Fierz-complete basis of four-fermion terms compatible with the symmetries of the Gross-Neveu-XY models consists of nine terms given by
\begin{align}\label{eq:Lintchi}
    \mathcal{L}_\mathrm{int} & = \frac{g_1}{2\Nf} (\bar{\psi} \psi)^2 + \frac{g_2}{2\Nf} (\bar{\psi} \gamma_\mu \psi)^2 + \frac{g_3}{2\Nf} (\bar{\psi} \gamma_2 \psi)^2 \nonumber\\ 
    &\quad
    + \frac{g_4}{2\Nf} [(\bar{\psi} \gamma_3 \psi)^2 + (\bar{\psi} \gamma_5 \psi)^2] + \frac{g_5}{2\Nf} (\bar{\psi} \gamma_{01} \psi)^2 \nonumber\\ 
    &\quad
    + \frac{g_6}{2\Nf} (\bar{\psi} \gamma_{\mu 2} \psi)^2 + \frac{g_7}{2\Nf} [(\bar{\psi} \gamma_{\mu3} \psi)^2 + (\bar{\psi} \gamma_{\mu5} \psi)^2] \nonumber\\
    &\quad
    + \frac{g_8}{2\Nf}[(\bar{\psi} \gamma_{23} \psi)^2 + (\bar{\psi} \gamma_{25} \psi)^2] + \frac{g_9}{2\Nf} (\bar{\psi}\gamma_{35}\psi)^2. 
\end{align}

\subsection{Flow equations}

We calculate the flow equations within the full Gross-Neveu-XY theory space given by the action
\begin{align}\label{eq:each}
    S & = \int \rmd^Dx \left( \bar{\psi}^\alpha \gamma_\mu \partial_\mu \psi^\alpha + \mathcal{L}_\mathrm{int} \right),
\end{align}
up to one-loop order in $D=2+\epsilon$ space-time dimensions, by employing the general one-loop formula derived in Ref.~\cite{Gehring:2015}. Evaluating the traces that appear in the formula, we obtain the beta functions
\begin{align}
    \beta_1 & = \epsilon g_1+\frac{1}{\Nf}[(2-4 \Nf) g_1^2+4 g_2 g_1+2 g_3 g_1 \nonumber\\*
    &\quad+4 g_4 g_1+2 g_5 g_1 +4 g_6 g_1+8 g_7 g_1+4 g_8 g_1 \nonumber\\*
    &\quad+2 g_9 g_1+4 g_2 g_5+4 g_3 g_6+8 g_4 g_7], \\
    \beta_2 & = \epsilon g_2+\frac{1}{\Nf}[2 g_1 g_5+4 g_4 g_8+2 g_3 g_9], \\
    \beta_3 & = \epsilon g_3+\frac{1}{\Nf}[(4 \Nf-2) g_3^2-2 g_1 g_3+4 g_2 g_3 \nonumber\\*
    &\quad+4 g_4 g_3-2 g_5 g_3 +4 g_6 g_3-8 g_7 g_3+4 g_8 g_3 \nonumber\\*
    &\quad-2 g_9 g_3+4 g_1 g_6+8 g_4 g_7+4 g_2 g_9], \\
    \beta_4 & = \epsilon g_4+\frac{1}{\Nf}[4 \Nf g_4^2-2 g_1 g_4+4 g_2 g_4+2 g_3 g_4 \nonumber\\*
    &\quad-2 g_5 g_4 +2 g_9 g_4+4 g_1 g_7+4 g_3 g_7+4 g_2 g_8], \\
    \beta_5 & = \epsilon g_5+\frac{1}{\Nf}[(4 \Nf-2) g_5^2-2 g_1 g_5+4 g_2 g_5 \nonumber\\* 
    &\quad-2 g_3 g_5-4 g_4 g_5 +4 g_6 g_5+8 g_7 g_5-4 g_8 g_5 \nonumber\\* 
    &\quad-2 g_9 g_5+4 g_1 g_2+8 g_7 g_8+4 g_6 g_9], \\
    \beta_6 & = \epsilon g_6+\frac{1}{\Nf}[2 g_4^2+8 g_7^2+2 g_8^2+2 g_1 g_3+2 g_5 g_9], \\
    \beta_7 & = \epsilon g_7+\frac{1}{\Nf}[2 g_1 g_4+2 g_3 g_4+8 g_6 g_7+2 g_5 g_8 \nonumber\\*
    &\quad+2 g_8 g_9], \\
    \beta_8 & = \epsilon g_8+\frac{1}{\Nf}[-4 \Nf g_8^2+2 g_1 g_8+4 g_2 g_8-2 g_3 g_8 \nonumber\\* 
    &\quad+2 g_5 g_8 -2 g_9 g_8+4 g_2 g_4+4 g_5 g_7+4 g_7 g_9], \\
    \beta_9 & = \epsilon g_9+\frac{1}{\Nf}[(2-4 \Nf) g_9^2+2 g_1 g_9+4 g_2 g_9 \nonumber\\* 
    &\quad+2 g_3 g_9-4 g_4 g_9 +2 g_5 g_9+4 g_6 g_9-8 g_7 g_9 \nonumber\\*
    &\quad-4 g_8 g_9+4 g_2 g_3+4 g_5 g_6+8 g_7 g_8],
\end{align}
where $\beta_i \equiv \partial_t g_i$ for the RG time $t$, $i=1,\dots,9$. Here, the sign of the beta functions is defined such that a coupling $g_i$ decreases (increases) in the flow towards the infrared if $\beta_i > 0$ ($\beta_i < 0$). 

We note that the above one-loop flow equations are invariant under the following exchanges of couplings
\begin{align}
    \mathcal{P}_1 & : && (g_1,g_2,g_3,g_4,g_5,g_6,g_7,g_8,g_9) \nonumber\\ &&&
    \mapsto (-g_5,g_2,-g_9,-g_8,-g_1,g_6,g_7,-g_4,-g_3), \\ 
    \mathcal{P}_2 & : && (g_1,g_2,g_3,g_4,g_5,g_6,g_7,g_8,g_9) \nonumber\\ &&&
    \mapsto (-g_3,g_2,-g_1,g_4,-g_9,g_6,-g_7,g_8,-g_5).
\end{align}
This property is useful to analyze the fixed-point structure in the nine-dimensional theory space, since if $g^\star$ is a fixed point, so is $\mathcal{P} g^\star$, and both share the same critical exponents at one-loop order.

\subsection{Fixed-point structure}

In the large-$\Nf$ limit, the flow equations decouple.
This allows us to identify the symmetry-breaking pattern associated with a given critical fixed point.
For instance, the fixed point located at 
\begin{align}
    g^\star_\mathrm{GNI} = [\Nf /(4 \Nf-2),0,0,0,0,0,0,0,0]\epsilon
\end{align}
corresponds to an instability at which the four-fermion coupling $g_1$ diverges at a finite RG time. This suggests that the fixed point describes Gross-Neveu-Ising criticality with $\mathbb{Z}_2$ order parameter $\langle \bar{\psi}\psi \rangle$, which breaks $\mathbb Z_2$ chiral symmetry~\cite{braun11}.
On the honeycomb lattice, it describes the quantum phase transition between the Dirac semimetal and a charge-density-wave insulator~\cite{HerbutGraphene:2006,HerbutInteracting:2009}. As no quadratic term $\propto g_1^2$ appears in the flow equations of the other couplings $g_2, \dots, g_9$, the Gross-Neveu-Ising model is closed under the RG, in agreement with previous results~\cite{Gehring:2015}. 
There are three additional Gross-Neveu-Ising-type fixed points, located at
\begin{align}
    g^\star_\mathrm{GNI'} &= [0,0,0,0,0,0,0,0,\Nf /(4 \Nf-2)]\epsilon, \\
    g^\star_\mathrm{GNI''} &= [0,0,-\Nf /(4 \Nf-2),0,0,0,0,0,0]\epsilon, \\
    g^\star_\mathrm{GNI'''} &= [0,0,0,0,-\Nf /(4 \Nf-2),0,0,0,0]\epsilon,
\end{align}
which describe Gross-Neveu-Ising-type criticality with order parameters 
$\langle \bar{\psi} \gamma_{35} \psi \rangle$, 
$\langle \bar{\psi}\rmi \gamma_2 \psi \rangle$, 
and 
$\langle \bar{\psi} \rmi \gamma_{01} \psi \rangle$,
respectively. 
These $\mathbb{Z}_2$ order parameters break 
time-reversal symmetry (GNI$'$), 
parity symmetry (GNI$''$),
and parity, time-reversal, and discrete chiral symmetry (GNI$'''$),
respectively.
GNI$'$ is the unique critical fixed point in the Gross-Neveu model using an irreducible representation of the Clifford algebra in $D=1+1$ or $D=2+1$ dimensions~\cite{gracey16}, and has previously been referred to as ``irreducible Gross-Neveu fixed point''~\cite{Gehring:2015}.
On the honeycomb lattice, GNI$'$ corresponds to a possible instability towards a quantum anomalous Hall insulator~\cite{raghu08, HerbutInteracting:2009}.
GNI$''$ and GNI$'''$ are specific to $D=1+1$ space-time dimensions and have no straightforward interpretation in $D=2+1$ dimensions.
Note that the fixed-point values of GNI (GNI$''$) map to those of GNI$'$ (GNI$'''$) under the combined transformation $\mathcal P_1 \mathcal P_2$.
Under $\mathcal P_1$, GNI maps to GNI$'''$ and GNI$'$ maps to GNI$''$.
At one-loop order, the critical behavior of all four Gross-Neveu-Ising-type fixed points are identical. This result is in line with the previous observation that the ``reducible'' GNI fixed point and the ``irreducible'' GNI$'$ fixed point feature the same exponents at order $\mathcal O(1/N)$ in the large-$N$ expansion in fixed $D=2+1$ dimensions~\cite{boyack19}. However, higher-order corrections within the $4-\epsilon$ expansion have been found to introduce slight differences in the universal critical behavior of the two fixed points, beginning at the fourth loop order~\cite{Zerf4L:2017}, see also the discussion in Ref.~\cite{Erramilli:2022kgp}.

Here, we are mainly interested in Gross-Neveu-XY criticality, described by the $\mathrm{U}(1)$ order parameter ${\langle \vec{\varphi} \rangle = \rmi (\langle \bar{\psi} \gamma_3 \psi \rangle, \langle \bar{\psi} \gamma_5 \psi \rangle)^\top}$. 
The fixed point naturally associated with this transition is located at
\begin{align} \label{eq:GNXY-FP}
    g^\star_\mathrm{GNXY} (\Nf > \Nfc) = [0,0,0,-1/4,0,-1/(8\Nf),0,0,0]\epsilon.
\end{align}
In the large-$\Nf$ limit, it lies on the $g_4$ axis and thus corresponds to an instability at which the O(2) vector $\vec\varphi$ develops a finite vacuum expectation value. By continuity, we assume that it governs the universal behavior of the Gross-Neveu-XY transition also at finite $\Nf$, as long as $\Nf$ is larger than a critical value $\Nfc$, to be discussed below.
A copy of this fixed point at one-loop order can be obtained by $\mathcal P_1$ and is located at
\begin{align} \label{eq:GNXYprime-FP}
    g^\star_\mathrm{GNXY'}(\Nf > \Nfc) = [0,0,0,0,0,-1/(8\Nf),0,1/4,0]\epsilon.
\end{align}
We demonstrate in Appendix~\ref{app:spinXY} that GNXY$'$ governs the critical behavior of the Gross-Neveu-Spin-XY model, as long as $\Nf > \Nfc$.
Due to the $\mathcal P_1$ symmetry of the flow equations, the original Gross-Neveu-XY model defined in Eq.~\eqref{eq:GNXYaction} and the Gross-Neveu-Spin-XY model thus share the same critical exponents at one-loop order, similar to the reducible and irreducible Gross-Neveu-Ising fixed points discussed above.
Whether this property also holds at higher loop orders represents an interesting direction for future work.

The flow equations of $g_i$ with $i \notin \{ 4,6\}$ do not contain terms $\propto g_j g_k$ with $j,k \in \{4,6\}$.
The subspace formed by $(g_4,g_6)$ is thus closed under the one-loop RG. We refer to this subspace as ``Gross-Neveu-XY subspace.''
Whether or not perturbations out of this subspace are relevant or irrelevant in the RG sense depends, in general, on the number of flavors $\Nf$, as well as the value of the couplings $g_4$ and $g_6$ within the subspace.
In the vicinity of a fixed point included in the Gross-Neveu-XY subspace, this question can be straightforwardly accessed by calculating the eigenvalues $\Theta_I$, $I=1,\dots,9$, and the corresponding eigendirections of the stability matrix $(- \partial \beta_i/\partial g_j)$ evaluated at the fixed point, with $i,j=1,\dots,9$. At one loop, there will be at least one eigenvalue $\Theta_1 = \epsilon$, which is a RG relevant direction lying fully within the subspace, and to which the correlation-length exponent $\nu$ is related by $\nu = 1/\Theta_1$~\cite{Gehring:2015, Ladovrechis:2023}. If all other eigendirections are RG irrelevant, i.e., $\Theta_I < 0$ for $I = 2,\dots,9$, the fixed point corresponds to a continuous quantum phase transition. In such case, we refer to it as a quantum critical fixed point.
For all $\Nf > 1$, we find that $\Theta_I < 0$ for $3 \leq I \leq 9$, but the sign of $\Theta_2$ turns out to depend on $\Nf$. In particular, we find $\Theta_2 = -(\Nf^2 - 2\Nf - 1)/\Nf^2\epsilon + \mathcal O(\epsilon^2)$ at one-loop order.
Hence, for $\Nf > \Nfc$ with
\begin{align}
\Nfc = 1+\sqrt{2} + \mathcal O(\epsilon) \simeq 2.41 + \mathcal O(\epsilon),
\end{align}
the Gross-Neveu-XY fixed point has a single RG relevant direction and represents a quantum critical fixed point.
For $\Nf \searrow \Nfc$, however, the Gross-Neveu-XY fixed point collides with another fixed point and exchanges stability with the latter for $\Nf < \Nfc$.
Without additional fine tuning, the Gross-Neveu-XY fixed point can no longer be accessed, rendering this fixed point unstable close to two dimensions.
Instead, a non-fine-tuned flow starting near the Gross-Neveu-XY axis parametrized by sizable $g_4 < 0$ and small perturbations $g_{1,3,7} < 0$ out of the Gross-Neveu-XY subspace is attracted at criticality by this other quantum critical fixed point.
For $\Nf = 2$, the other quantum critical fixed point is located at
\begin{align} \label{eq:GNXY-FP-Nf=2}
    g^\star_\mathrm{GNXY}(\Nf = 2) = \bigg[\!-\frac{1}{6},0,\!-\frac{1}{6},\!-\frac{1}{6},0,\!-\frac{1}{12},\!-\frac{1}{12},0,0\bigg] \epsilon.
\end{align}
Assuming that the runaway flow associated with the latter fixed point still corresponds to an instability at which the O(2) vector $\vec\varphi$ develops a finite vacuum expectation value, it is then the latter fixed point that governs the universal behavior of the Gross-Neveu-XY transition for $\Nf < \Nfc$.
There are three additional copies of this fixed point that can be obtained by applying $\mathcal{P}_1$, $\mathcal{P}_2$ and $\mathcal{P}_1 \mathcal{P}_2$,
\begin{align} 
    g^\star_\mathrm{GNXY'}(\Nf = 2) &= \bigg[0,0,0,0,\frac{1}{6},-\frac{1}{12},-\frac{1}{12},\frac{1}{6},\frac{1}{6}\bigg] \epsilon, \\
    g^\star_\mathrm{GNXY''}(\Nf = 2) &= \bigg[\frac{1}{6},0,\frac{1}{6},-\frac{1}{6},0,-\frac{1}{12},\frac{1}{12},0,0\bigg] \epsilon, \\
    g^\star_\mathrm{GNXY'''}(\Nf = 2) &= \bigg[0,0,0,0,-\frac{1}{6},-\frac{1}{12},\frac{1}{12},\frac{1}{6},-\frac{1}{6}\bigg] \epsilon.
\end{align}
They govern the non-fine-tuned flow starting near the Gross-Neveu-XY axis for other types of perturbations, e.g., small $g_{1,3,7} > 0$ in the case of GNXY$''$.
At one-loop order, all of these fixed points feature the same critical behavior as the fixed point given in Eq.~\eqref{eq:GNXY-FP-Nf=2}.

%%%%%%%%%%%%%%%%%%%%%%%%%%%%%%%%%%%%%%%%%%%%%%%%%%%%%%%%%%%%%%%%%%%%%%%%%%%%%%%%
\section{Quantum critical behavior}\label{sec:critexp}
%%%%%%%%%%%%%%%%%%%%%%%%%%%%%%%%%%%%%%%%%%%%%%%%%%%%%%%%%%%%%%%%%%%%%%%%%%%%%%%%

We characterize the quantum critical behavior of the Gross-Neveu-XY model in terms of the universal exponents $1/\nu$, $\eta_\varphi$, and $\eta_\psi$. The remaining exponents can be derived from hyperscaling relations~\cite{Herbut:2007}.
We start with the calculation in $2+\epsilon$ space-time dimensions and then combine the expansion results near the lower and the upper critical dimensions to obtain improved estimates in the physical dimension using Pad\'e interpolation.

\subsection{\(\boldsymbol{2+\epsilon}\) expansion} \label{ssec:2pepsresults}

\subsubsection{Correlation-length exponent}
A quantum critical fixed point is characterized by a unique positive eigenvalue $\Theta_1 > 0$ of the stability matrix $(-\partial \beta_i / \partial g_j)$.
The exponent $\nu$ that governs the divergence of the correlation length $\xi \propto |g - g^\star|^{-\nu}$, where $|g - g^\star|$ corresponds to the distance to the critical point, is given by $\nu  = 1/\Theta_1$.
At the Gross-Neveu-XY fixed point for $\Nf > \Nfc$, as well as at the critical fixed point governing the Gross-Neveu-XY criticality for $\Nf < \Nfc$, we find
\begin{align} \label{eq:nu}
    1/\nu = \epsilon + \mathcal{O}(\epsilon^2),
\end{align} 
independent of $\Nf$, in agreement with the general result known for Gross-Neveu-type models at one-loop order~\cite{Gehring:2015, Ladovrechis:2023}.
The above equation agrees also with the large-$\Nf$ result~\cite{GraceyN3XY:2021}, when expanding the latter in small $\epsilon = D-2$,%
\footnote{Note that the parameters $N$ and $N_t$ appearing in Ref.~\cite{GraceyN3XY:2021} are related to the flavor number $\Nf$ of this work by $N = 4N_t = 4 \Nf$.}
\begin{align}
    1/\nu = \epsilon - \frac{1}{2\Nf^2} \epsilon^2 + \mathcal O(1/\Nf^3,\epsilon^3).
\end{align}
Our result in Eq.~\eqref{eq:nu} shows that the zeroth-order large-$\Nf$ result becomes exact for all $\Nf$ to leading order in $\epsilon = D - 2$.

\subsubsection{Order-parameter anomalous dimension}

At the quantum critical point, the dynamical structure factor $\mathcal S$ as function of momentum $\mathbf k$ and real frequency $\omega$ is characterized by a power-law form 
$\mathcal S(\mathbf k,\omega) \propto 1/(\omega^2 - \mathbf k^2)^{(2 - \eta_\varphi)/2}$ with anomalous dimension $\eta_\varphi$.
In order to compute $\eta_\varphi$, we add an infinitesimal symmetry-breaking term to the fixed-point action
\begin{align}
    S \mapsto S + y \int \rmd^Dx\, \rmi \bar{\psi} \mathcal{M} \psi
\end{align}
with small $y \in \mathbbm R$ and $\mathcal{M} \in \{\gamma_3, \gamma_5\}$. Assuming hyperscaling to hold, $\eta_\varphi$ can be obtained from the flow of the parameter $y$~\cite{JanssenLorentz:2016,Ladovrechis:2023}. Specifically, the anomalous dimension $\eta_\varphi$ is related to the scaling dimension $\Delta_y$ of $y$ at the quantum critical fixed point by $\eta_\varphi = D + 2(1-\Delta_y)$.
The scaling dimension of $y$ has previously been computed at one-loop order for a general relativistic four-fermion theory, leading to the general formula for the anomalous dimension~\cite{Ladovrechis:2023}
\begin{align}
    \eta_\varphi = 2 + \epsilon - 2\sum_i c_i g_i^\star,
\end{align}
where $g^\star_i$, $i=1,\dots,9$, denotes the fixed-point couplings and the coefficients $c_i$ are given as
\begin{align}
c_i &= \frac{1}{4D\Nf} \sum_\mu [\Nf \mathrm{Tr}(\mathcal{M} \gamma_\mu O_i \gamma_\mu) \mathrm{Tr}(\mathcal{M} O_i)
\nonumber \\ & \quad
- \mathrm{Tr}(O_i \gamma_\mu \mathcal{M} \gamma_\mu O_i \mathcal{M})] \quad \text{(no sum over $i$)},
\end{align}
where $D$ corresponds to the space-time dimension.
Evaluating the traces, we find
\begin{align} \label{eq:eta-phi}
\eta_\varphi(\Nf > \Nfc) &= 2 - \left(1-\frac{1}{2\Nf^2}\right) \epsilon + \mathcal{O}(\epsilon^2)
\end{align}
at the Gross-Neveu-XY fixed point for $\Nf > \Nfc$ [Eq.~\eqref{eq:GNXY-FP}].
At the quantum critical fixed point governing the universal behavior of the Gross-Neveu-XY transition for $\Nf < \Nfc$ [Eq.~\eqref{eq:GNXY-FP-Nf=2}], we find
\begin{align} \label{eq:eta-phi-Nf=2}
    \eta_\varphi(\Nf = 2) = 2 - \frac{1}{6} \epsilon + \mathcal O(\epsilon^2)
\end{align}
for $\Nf = 2$.
Equation~\eqref{eq:eta-phi} agrees with the large-$\Nf$ result~\cite{GraceyN3XY:2021}, when expanding the latter in small $\epsilon = D-2$,
\begin{align}
\eta_\varphi = 2 - \left( 1 - \frac{1}{2\Nf^2} \right) \epsilon - \frac{4(1+\Nf)}{8 \Nf^2} \epsilon^2 + \mathcal{O}(1/\Nf^3,\epsilon^3).
\end{align}
Our result in Eq.~\eqref{eq:eta-phi} shows that the second-order large-$\Nf$ result becomes exact for all $\Nf > \Nfc$ to leading order in $\epsilon = D-2$.
The fixed-point collision occurring as a function of $\Nf$ at $\Nf = \Nfc$, however, cannot be perturbatively accessed through an expansion in powers of $1/\Nf$.
Thus, Eq.~\eqref{eq:eta-phi-Nf=2} for $\Nf = 2$ represents a result that goes beyond the large-$\Nf$ expansion.

\subsubsection{Fermionic anomalous dimension}

At the quantum critical point, the fermion spectral function $A$ has also a power-law form $A(\mathbf k, \omega) \propto 1/(\omega^2 - \mathbf k^2)^{(1-\eta_\psi)/2}$ with fermion anomalous dimension $\eta_\psi$.
At one-loop order near the lower critical dimension, $\eta_\psi$ vanishes, such that the first non-trivial contribution arises only at two-loop order~\cite{Gehring:2015, JanssenThirring:2010}.
A general formula for $\eta_\psi$ at two-loop order for relativistic four-fermion theories has been computed in Ref.~\cite{Ladovrechis:2023},
\begin{align}
    \eta_\psi &= \sum_{i,j} g_i^\star H_{ij} g_j^\star,
\end{align}
with the coefficients $H_{ij}$ given by
\begin{align}
H_{ij} &= \frac{1}{32 \Nf^2} \sum_{\mu,\nu,\lambda} (\delta_{\mu 0}\delta_{\nu \lambda} + \delta_{\nu 0}\delta_{\mu \lambda} + \delta_{\lambda 0}\delta_{\mu \nu})
\nonumber \\&\quad
\times \bigl[\Nf \mathrm{Tr}(\gamma_0 O_i \gamma_\mu O_j) \mathrm{Tr}(\gamma_\nu O_j \gamma_\lambda O_i)
\nonumber \\&\quad
-\mathrm{Tr}(\gamma_0 O_i \gamma_\mu O_j \gamma_\nu O_i \gamma_\lambda O_j)\bigr]
\quad \text{(no sum over $i,j$)}.
\end{align}
Evaluating the traces, we find
\begin{align} \label{eq:eta-psi}
\eta_\psi(\Nf > \Nfc) &= \frac{4\Nf^2 - 1}{16\Nf^3} \epsilon^2 + \mathcal{O}(\epsilon^3).
\end{align}
at the Gross-Neveu-XY fixed point for $\Nf > \Nfc$ [Eq.~\eqref{eq:GNXY-FP}].
At the quantum critical fixed point governing the universal behavior of the Gross-Neveu-XY transition for $\Nf < \Nfc$ [Eq.~\eqref{eq:GNXY-FP-Nf=2}], we find
\begin{align} \label{eq:eta-psi-Nf=2}
\eta_\psi(\Nf = 2) = \frac{7}{72} \epsilon^2 + \mathcal O(\epsilon^3).
\end{align}
for $\Nf = 2$.
Equation~\eqref{eq:eta-psi} agrees with the large-$\Nf$ result~\cite{GraceyN3XY:2021}, when expanding the latter in small $\epsilon = D-2$,
\begin{align}
\eta_\psi = \frac{4 \Nf^2 - 1}{16 \Nf^3} \epsilon^2 + \frac{1 - \Nf(\Nf - 2)}{8 \Nf^3} \epsilon^3 + \mathcal{O}(1/\Nf^4,\epsilon^4).
\end{align}
Our result in Eq.~\eqref{eq:eta-psi} shows that the third-order large-$\Nf$ result becomes exact for all $\Nf > \Nfc$ to leading order in $\epsilon = D-2$.
Equation~\eqref{eq:eta-psi-Nf=2} for $\Nf = 2$, by contrast, which arises as a consequence of the fixed-point collision as a nonperturbative phenomenon in $1/\Nf$, goes beyond the large-$\Nf$ expansion.

%%%%%%%%%%%%%%%%%%%%%%%%%%%%%%%%%%%%%%%%%%%%%%%%%%%%%%%%%%%%%%%%%%%%%%%%%%%%%%%%
\begin{figure*}[tb!]
\includegraphics[width=\textwidth]{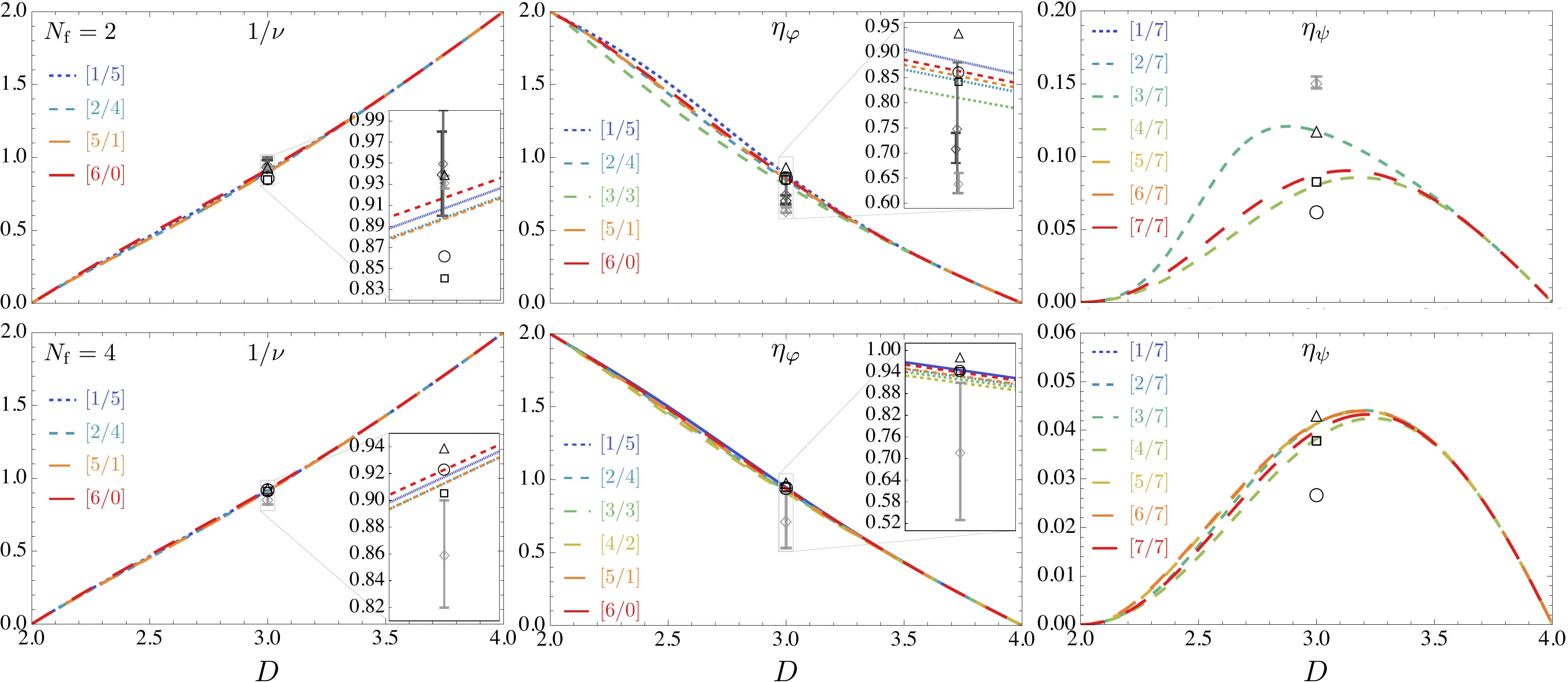}
\caption{%
Critical exponents $1/\nu$ (left column), $\eta_\varphi$ (center column), and $\eta_\psi$ (right column) for $\Nf = 2$ (top row) and $\Nf = 4$ (bottom row) as functions of space-time dimension $D$ from Pad\'e interpolation $[m/n]$ between expansions near lower (Sec.~\ref{ssec:2pepsresults}) and upper (Ref.~\cite{Zerf4L:2017}) critical dimensions, in comparison with previous results from large-$\Nf$ expansion (squares)~\cite{GraceyN3XY:2021}, functional RG (circles)~\cite{ClassenXY:2017}, $4-\varepsilon$ expansion (triangles)~\cite{Zerf4L:2017}, and quantum Monte Carlo simulations (diamonds)~\cite{LiFermi:2017, Xu:2018, Otsuka:2018, Huang:2024}.
For better visibility, the quantum Monte Carlo results have been minimally shifted towards the left.
}
\label{fig:crits}
\end{figure*}
%%%%%%%%%%%%%%%%%%%%%%%%%%%%%%%%%%%%%%%%%%%%%%%%%%%%%%%%%%%%%%%%%%%%%%%%%%%%%%%%

\subsubsection{Corrections-to-scaling exponent}

In the direct vicinity of the quantum critical point, various observables feature power laws arising from the scale invariance emerging at criticality.
Moving further away from the quantum critical point, corrections to scaling appear, for instance, in the correlation length $\xi \propto | g - g^\star|^{-\nu}(1 + A | g - g^\star|^\omega + \dots)$, with nonuniversal coefficient $A$ and universal exponent $\omega$.
In the RG approach, the corrections-to-scaling-exponent is given by the negative of the second-largest eigenvalue $\Theta_2 < 0$ of the stability matrix $(-\partial \beta_i / \partial g_j)$ at the corresponding critical fixed point, $\omega = - \Theta_2$.
At the Gross-Neveu-XY fixed point for $\Nf > \Nfc$ [Eq.~\eqref{eq:GNXY-FP}], we find
\begin{align}\label{eq:corr2scal}
\omega(\Nf > \Nfc) = \frac{\Nf^2 - 2\Nf - 1}{\Nf^2} \epsilon + \mathcal{O}(\epsilon^2),
\end{align}
while for the quantum critical fixed point governing the universal behavior of the Gross-Neveu-XY transition for $\Nf < \Nfc$ [Eq.~\eqref{eq:GNXY-FP-Nf=2}], we find
\begin{align}
\omega(\Nf = 2) = \frac{1}{3} \epsilon + \mathcal{O}(\epsilon^2).
\end{align}
The above equations originate from the flow in the Fierz-complete basis of four-fermion terms and, therefore, cannot be meaningfully compared with results from Fierz-incomplete approaches, such as conventional large-$\Nf$ or $4-\varepsilon$ expansions.

\subsection{Interpolation of \(\boldsymbol{2+\epsilon}\) and \(\boldsymbol{4-\varepsilon}\) expansions}
\label{subsec:critexpint}

The purely fermionic action in Eq.~\eqref{eq:GNXYaction} can be partially bosonized by means of a Hubbard-Stratonovich transformation. The resulting Yukawa coupling $h$ has mass dimension $[h] = 4-D$, and thus admits an expansion around the upper critical dimension $D_\mathrm{c}^\mathrm{up} = 4$. The previous four-loop RG analysis~\cite{Zerf4L:2017} has provided estimates for the critical exponents $1/\nu$, $\eta_\varphi$, and $\eta_\psi$ up to fourth order in $\varepsilon = 4 - D$.
Extrapolating towards the physical dimension $D = 2+1$, however, requires an appropriate resummation of the different terms.
For the Gross-Neveu-Ising~\cite{JanssenAFM:2014, IhrigIsing:2018} and Gross-Neveu-Heisenberg~\cite{Ladovrechis:2023} cases, it has been shown that estimates can be significantly improved by employing a suitable interpolation between extrapolation results from both the upper and lower critical dimensions, rather than relying solely on extrapolation from one side.
In this section, we combine our results from the $2+\epsilon$ expansion (Sec.~\ref{ssec:2pepsresults}) with the results from the $4-\varepsilon$ expansion (Ref.~\cite{Zerf4L:2017}) to obtain improved estimates for the critical exponents characterizing the Gross-Neveu-XY universality class in the physical space-time dimension $D=2+1$.

Following previous works~\cite{JanssenAFM:2014, IhrigIsing:2018, Ladovrechis:2023}, we employ an interpolation scheme based on Padé approximants $[m/n]$, defined as
\begin{align}
    [m/n](D) = \frac{\sum_{i=0}^m a_i D^i}{1 + \sum_{j=1}^n b_j D^j},
\end{align}
with $m,n \in \mathbbm N_0$, and $m+n+1$ coefficients $a_i$ and $b_j$ chosen to ensure that $[m/n](2+\epsilon)$ reproduces our results from Sec.~\ref{ssec:2pepsresults} up to the provided order in $\epsilon$, while $[m/n](4-\varepsilon)$ matches the four-loop results of Ref.~\cite{Zerf4L:2017}. We refer to this approach as Padé interpolation.
The requirement that the interpolation agrees with both the upper and lower critical dimensions imposes a finite set of constraints that must be satisfied. To ensure a unique solution, the number of coefficients, $m+n+1$, is chosen to match the number of constraints.
In particular, our determination of $1/\nu$ and $\eta_\varphi$ to first order in $\epsilon = D-2$, combined with the previous fourth-order determination in $\varepsilon = 4-D$, imposes seven constraints, leading to $m+n+1 = 7$.
For $\eta_\psi$, which we have determined to second order in $\epsilon$, we have one additional constraint, resulting in $m+n+1=8$.
This implies that a number of different Padé approximants may be used, and it is a priori not clear which one will give the most faithful estimate for the physical dimension $D = 2+1$. However, some choices of $m$ and $n$ can be excluded, if
(1)~singularities appear for $D \in (2,4)$, or
(2)~not all constraints can be simultaneously fulfilled.
The latter is the case for $1/\nu$ and $\eta_\varphi$ if $m = 0$, and for $\eta_\psi$ if $m=0,1,2$.

Figure~\ref{fig:crits} presents our results for the admissible Padé interpolations of $1/\nu$, $\eta_\varphi$, and $\eta_\psi$ as functions of $D$ for $\Nf = 2$ and $\Nf = 4$.
The corresponding numerical values in $D=2+1$ space-time dimensions are given in Tables~\ref{tab:Nf4crits} and \ref{tab:Nf2crits}.
Our best estimate is obtained by averaging over the corresponding non-singular Padé approximants in each case, given in boldface font in the respective table.
These results are compared with estimates from the large-$\Nf$ expansion~\cite{GraceyN3XY:2021}, functional RG~\cite{ClassenXY:2017}, and quantum Monte Carlo simulations~\cite{LiFermi:2017, Otsuka:2018, Xu:2018, Huang:2024} in Tables~\ref{tab:summaryNf=4} and \ref{tab:summaryNf=2}.
For an unbiased comparison with results from the $4-\varepsilon$ and large-$\Nf$ expansions, we present the averaged results over all admissible Padé approximants, following the same criteria defined above.
The individual Pad\'e approximants are given in Appendix~\ref{app:additional-data}.

%%%%%%%%%%%%%%%%%%%%%%%%%%%%%%%%%%%%%%%%%%%%%%%%%%%%%%%%%%%%%%%%%%%%%%%%%%%%%%%%
\begin{table}[tb!]
\caption{%
Critical exponents of the Gross-Neveu-XY fixed point in $D=2+1$ space-time dimensions for $\Nf = 4$ four-component Dirac fermions from different Padé interpolations $[m/n]$ between first-order (top) and second-order (bottom) expansions near the lower critical dimension (Sec.~\ref{ssec:2pepsresults}) and fourth-order expansion near the upper critical dimension (Ref.~\cite{Zerf4L:2017}). 
Non-admissible Padé approximants, which exhibit singularities for $D \in (2,4)$ or are non-existent due to failing to satisfy all constraints, are marked as ``sing.'' and ``n.e.,'' respectively.
The dashes ``--'' signify approximants for which the required $\epsilon^2$ corrections are not yet available.
The numbers in boldface represent the averages (``avg.'') and deviations of the admissible Pad\'e interpolations.
}
\begin{tabular*}{\linewidth}{@{\extracolsep{\fill}} l c l l l}
\noalign{\smallskip}\hline\hline\noalign{\smallskip}
 $\Nf = 4$ & $[m/n]$ & \multicolumn{1}{c}{$1/\nu$} & \multicolumn{1}{c}{$\eta_\varphi$} & \multicolumn{1}{c}{$\eta_\psi$} \\
\noalign{\smallskip}\hline\noalign{\smallskip}
 $\mathcal{O}(\epsilon,\varepsilon^4)$ & [1/5] & 0.917395 & 0.945021 & \multicolumn{1}{c}{n.e.}\\
 & [2/4] & 0.912634 & 0.926743 & \multicolumn{1}{c}{n.e.}\\
 & [3/3] & \multicolumn{1}{c}{sing.} & 0.919012 & \multicolumn{1}{c}{sing.}\\
 & [4/2] & \multicolumn{1}{c}{sing.} & 0.909690 & 0.0416190 \\
 & [5/1] & 0.912682 & 0.928530 & 0.0416928 \\
 & [6/0] & 0.922746 & 0.938542 & 0.0416501 \\
 & \textbf{avg.} & \textbf{0.916(5)} & \textbf{0.926(13)} & \textbf{0.04165(4)}
 \\
\noalign{\smallskip}
 $\mathcal{O}(\epsilon^2,\varepsilon^4)$ & [3/4] & \multicolumn{1}{c}{--} & \multicolumn{1}{c}{--} & 0.0413287\\
  & [4/3] & \multicolumn{1}{c}{--} & \multicolumn{1}{c}{--} & 0.0382788\\
  & [5/2] & \multicolumn{1}{c}{--} & \multicolumn{1}{c}{--} & 0.0413572\\
  & [6/1] & \multicolumn{1}{c}{--} & \multicolumn{1}{c}{--} & 0.0414591\\
  & [7/0] & \multicolumn{1}{c}{--} & \multicolumn{1}{c}{--} & 0.0399013\\ 
  & \textbf{avg.} & \multicolumn{1}{c}{--} & \multicolumn{1}{c}{--} & \textbf{0.0404(13)} \\
\noalign{\smallskip}\hline\hline
\end{tabular*}
\label{tab:Nf4crits}
\end{table}
%%%%%%%%%%%%%%%%%%%%%%%%%%%%%%%%%%%%%%%%%%%%%%%%%%%%%%%%%%%%%%%%%%%%%%%%%%%%%%%%

In the above results, for the expansion near the lower critical dimension, the exponents of the Gross-Neveu-XY fixed point in Eq.~\eqref{eq:GNXY-FP} have been used for both $\Nf = 2$ and $\Nf = 4$, despite this fixed point developing a second relevant direction for $\Nf < \Nfc$.
This choice is motivated by the fact that this fixed point constitutes the proper continuation of the Gross-Neveu-XY fixed point previously studied in the $4-\varepsilon$ expansion~\cite{Zerf4L:2017}.
As a consequence, the $\Nf = 2$ results presented in Fig.~\ref{fig:crits} and Table~\ref{tab:Nf2crits} come with an important caveat: If $\Nfc > 2$ in $D = 2+1$, as is the case near the lower critical dimension, the presented exponents are accessible in a lattice model only with additional fine-tuning. A generic Gross-Neveu-XY quantum phase transition, such as those studied in quantum Monte Carlo simulations in Refs.~\cite{LiFermi:2017, Xu:2018, Otsuka:2018}, would instead fall into the universality class defined by the quantum critical fixed point in Eq.~\eqref{eq:GNXY-FP-Nf=2}, with exponents differing from those of the Gross-Neveu-XY fixed point in Eq.~\eqref{eq:GNXY-FP}.
Since $\Nfc = 1 + \sqrt{2} + \mathcal{O}(\epsilon)$ is only slightly above $\Nf = 2$ near the lower critical dimension, higher-order terms are necessary to reliably determine which of the two fixed points defines the Gross-Neveu-XY universality class for $\Nf = 2$ four-component Dirac fermions in $D = 2+1$ space-time dimensions. This question is left for the future.

%%%%%%%%%%%%%%%%%%%%%%%%%%%%%%%%%%%%%%%%%%%%%%%%%%%%%%%%%%%%%%%%%%%%%%%%%%%%%%%%
\begin{table}[tb!]
\caption{%
Same as Table~\ref{tab:Nf4crits}, but for $\Nf = 2$ four-component Dirac fermions.
Note that the Gross-Neveu-XY fixed point becomes unstable for $\Nf < \Nfc$, with $\Nfc = 1 + \sqrt{2} + \mathcal O(\epsilon)$. If $\Nfc > 2$ in $D = 2+1$ space-time dimensions, the presented exponents are accessible in a lattice simulation only with fine-tuning.
}
\begin{tabular*}{\linewidth}{@{\extracolsep{\fill}} l c l l l}
\noalign{\smallskip}\hline\hline\noalign{\smallskip}
 $\Nf = 2$ & $[m/n]$ & \multicolumn{1}{c}{$1/\nu$} & $\eta_\varphi$ & $\eta_\psi$ \\
\noalign{\smallskip}\hline\noalign{\smallskip}
 $\mathcal{O}(\epsilon,\varepsilon^4)$ & [1/5] & 0.907038  & 0.882504 & \multicolumn{1}{c}{n.e.} \\
 & [2/4] & 0.898119 & 0.844309 & \multicolumn{1}{c}{n.e.} \\
 & [3/3] & \multicolumn{1}{c}{sing.} & 0.809482 & \multicolumn{1}{c}{sing.} \\
 & [4/2] & \multicolumn{1}{c}{sing.} & \multicolumn{1}{c}{n.e.} & \multicolumn{1}{c}{sing.} \\
 & [5/1] & 0.896979 & 0.853247 & \multicolumn{1}{c}{sing.} \\
 & [6/0] & 0.916586 & 0.862263 & 0.0996709 \\
 & \textbf{avg.} & \textbf{0.904(9)} & \textbf{0.850(27)} & \textbf{0.010}\\
\noalign{\smallskip}
$\mathcal{O}(\epsilon^2,\varepsilon^4)$ & [3/4] & \multicolumn{1}{c}{--} & \multicolumn{1}{c}{--} & 0.117487 \\
  & [4/3] & \multicolumn{1}{c}{--} & \multicolumn{1}{c}{--} & 0.0803001 \\
  & [5/2] & \multicolumn{1}{c}{--} & \multicolumn{1}{c}{--} & \multicolumn{1}{c}{sing.}\\
  & [6/1] & \multicolumn{1}{c}{--} & \multicolumn{1}{c}{--} & \multicolumn{1}{c}{sing.}\\
  & [7/0] & \multicolumn{1}{c}{--} & \multicolumn{1}{c}{--} & 0.087469 \\ 
  & \textbf{avg.} &\multicolumn{1}{c}{--} & \multicolumn{1}{c}{--} & \textbf{0.095(19)} \\
\noalign{\smallskip}\hline\hline
\end{tabular*}
\label{tab:Nf2crits}
\end{table}
%%%%%%%%%%%%%%%%%%%%%%%%%%%%%%%%%%%%%%%%%%%%%%%%%%%%%%%%%%%%%%%%%%%%%%%%%%%%%%%%

%%%%%%%%%%%%%%%%%%%%%%%%%%%%%%%%%%%%%%%%%%%%%%%%%%%%%%%%%%%%%%%%%%%%%%%%%%%%%%%%
\section{Conclusion}
\label{sec:conclusions}
%%%%%%%%%%%%%%%%%%%%%%%%%%%%%%%%%%%%%%%%%%%%%%%%%%%%%%%%%%%%%%%%%%%%%%%%%%%%%%%%

In this work, we have mapped out the RG fixed-point structure of the Gross-Neveu-XY theory space, relevant to various quantum phase transitions in the context of graphene and moir\'e Dirac materials.
To this end, we have examined purely fermionic representatives of the Gross-Neveu-XY universality class, which allow for a systematic expansion around the lower critical space-time dimension $D_\mathrm{c}^\mathrm{low} = 2$.
Since generically all symmetry-allowed operators are generated by a RG transformation, we have constructed a Fierz-complete basis of four-fermion operators invariant under the symmetries of the Gross-Neveu-XY model. The space spanned by this basis defines the Gross-Neveu-XY theory space relevant for the $2 + \epsilon$ expansion.
We have carried out a one-loop RG calculation in the Gross-Neveu-XY theory space to identify the Gross-Neveu-XY fixed point.
Close to two space-time dimensions, we have found that the Gross-Neveu-XY fixed point is stable only if $\Nf > \Nfc = 1 + \sqrt{2} + \mathcal{O}(\epsilon)$.
For $\Nf < \Nfc$, and without additional fine-tuning, a lattice model realizing a Gross-Neveu-XY quantum phase transition falls into the universality class defined by a different quantum critical fixed point, with exponents differing from
those of the (now unstable) Gross-Neveu-XY fixed point.
This scenario may be relevant for the semimetal-to-superconductor transition on the triangular lattice~\cite{Otsuka:2018} and the semimetal-to-Kekulé-insulator transition on the honeycomb lattice~\cite{LiFermi:2017, Xu:2018}, and potentially also for the recent experiments on tetralayer WSe$_2$~\cite{ma24}. For these cases, the number of four-component fermions is $\Nf = 2$.
To determine whether $\Nf = 2$ lies above or below $\Nfc$ in $D = 2+1$ space-time dimensions, future work should focus on computing the scaling dimensions of the symmetry-allowed four-fermion terms that become relevant at the Gross-Neveu-XY fixed point as the space-time dimension approaches $D_\mathrm{c}^\mathrm{low}$. This includes the terms parametrized by the couplings $g_1$, $g_3$, and $g_7$ in Eq.~\eqref{eq:Lintchi}. In addition to a higher-order $2+\epsilon$ expansion, this analysis could be conducted using the $4-\varepsilon$ expansion~\cite{dipietro16}, the large-$\Nf$ approach~\cite{benvenuti19}, or by studying the functional RG flow within the dynamical bosonization framework~\cite{gies02, Pawlowski:2005xe, janssen17, moser24}.
In contrast, the case of $\Nf = 4$, which is relevant for the twist-tuned transition in moiré bilayer graphene~\cite{Biedermann:2024, Huang:2024}, appears to be sufficiently above $\Nfc$ in the expansion around the lower critical dimension. As a result, we expect the transition in the physical space-time dimension $D = 2+1$ to be governed by the Gross-Neveu-XY fixed point.

For both the Gross-Neveu-XY fixed point and the quantum critical fixed point for $\Nf < \Nfc$, we have calculated the correlation-length exponent $1/\nu$, the anomalous dimension $\eta_\varphi$, and the corrections-to-scaling exponent $\omega$ to leading order in $\epsilon$, and the fermion anomalous dimension $\eta_\psi$ to second order in $\epsilon$.
For the Gross-Neveu-XY fixed point, we have furthermore obtained improved estimates for the exponents in $D=2+1$ space-time dimensions by using a Pad\'e approximant interpolation scheme, bridging our results with previous four-loop calculations from the expansion around the upper critical dimension~\cite{Zerf4L:2017}.
Our results are presented alongside comparisons with previous estimates in Tables~\ref{tab:summaryNf=4} and \ref{tab:summaryNf=2}.

%%%%%%%%%%%%%%%%%%%%%%%%%%%%%%%%%%%%%%%%%%%%%%%%%%%%%%%%%%%%%%%%%%%%%%%%%%%%%%%%
\begin{acknowledgments}
We thank Jan Biedermann, David Moser, and Mireia Tolosa-Sime\'on for discussions and collaboration on related topics. BH and MMS are supported by the Mercator Research Center Ruhr under Project No.~Ko-2022-0012.
MMS acknowledges funding from the Deutsche Forschungsgemeinschaft (DFG, German Research Foundation) under Project No.~277146847 (SFB 1238, project C02) and Project No.~452976698 (Heisenberg program).
The work of LJ is funded by the DFG under 
Project No.~247310070 (SFB 1143, Project A07), 
Project No.~390858490 (W\"urzburg-Dresden Cluster of Excellence {\it ct.qmat}, EXC 2147), and 
Project No.~411750675 (Emmy Noether program, JA2306/4-1).
\end{acknowledgments}
%%%%%%%%%%%%%%%%%%%%%%%%%%%%%%%%%%%%%%%%%%%%%%%%%%%%%%%%%%%%%%%%%%%%%%%%%%%%%%%%

\appendix

%%%%%%%%%%%%%%%%%%%%%%%%%%%%%%%%%%%%%%%%%%%%%%%%%%%%%%%%%%%%%%%%%%%%%%%%%%%%%%%%
\section{Gross-Neveu-Spin-XY Model}\label{app:spinXY}
%%%%%%%%%%%%%%%%%%%%%%%%%%%%%%%%%%%%%%%%%%%%%%%%%%%%%%%%%%%%%%%%%%%%%%%%%%%%%%%%

In this appendix, we discuss an alternative model that shares the $\mathrm{U}(\Nf) \times \mathrm{U}(\Nf)$ global symmetry with the Gross-Neveu-XY model defined in Eq.~\eqref{eq:GNXYaction}.
It is defined by the action $S = \int \rmd^Dx\, \mathcal{L}$ with
\begin{align}
    \mathcal{L} & =  \bar{\psi}^\alpha (\Gamma_\mu \otimes \mathbb{1}_2) \partial_\mu \psi^\alpha
\nonumber \\* & \quad
     + \frac{g}{2 \Nf} \bigl\{
    [\bar{\psi}^\alpha (\mathbb{1}_2 \otimes \sigma_x) \psi^\alpha]^2
    +[\bar{\psi}^\alpha (\mathbb{1}_2 \otimes \sigma_y) \psi^\alpha]^2 \bigr\}\,.
\end{align}
Here, $\psi^\alpha$ represents a four-component spinor composed of two Dirac fermion flavors corresponding to the two different spin polarizations.
The $2 \times 2$ matrices $\Gamma_\mu$ form an irreducible two-dimensional representation of the Clifford algebra. Similar to the above, $\alpha = 1,\dots,\Nf$ denotes the flavor index and ${\mu = 0,\dots,D-1}$ the space-time index. 

The Gross-Neveu-Spin-XY model preserves the same relativistic, flavor, and charge conservation symmetries as the Gross-Neveu-XY model in Eq.~\eqref{eq:GNXYaction}.
However, since it is defined using an irreducible representation of the Clifford algebra in $2 < D < 4$, it does not exhibit any continuous chiral symmetry generated by combinations of $\Gamma_\mu$.
Instead, it features a $\mathrm{U}(1)$ spin rotational symmetry generated by the Pauli matrix $\sigma_z$, under which four-component spinors transform as
\begin{align}
\psi^\alpha \mapsto \rme^{\rmi \phi \cdot (\mathbb{1}_2 \otimes \sigma_z)} \psi^\alpha
\end{align}
with rotation angle $\phi$.
The Gross-Neveu-Spin-XY model can be understood as deformation of the Gross-Neveu-Heisenberg model defined in Eq.~\cite{Ladovrechis:2023}, in which the four-fermion interaction term with $\vec \sigma = (\sigma_x, \sigma_y, \sigma_z)$ is replaced by
\begin{align}
    [\bar{\psi} (\mathbb{1}_2 \otimes \vec\sigma) \psi]^2 & \mapsto 
    [\bar{\psi} (\mathbb{1}_2 \otimes \sigma_{x}) \psi]^2
    + [\bar{\psi} (\mathbb{1}_2 \otimes \sigma_{y}) \psi]^2.
\end{align}
The deformation reduces the SU(2) spin symmetry to a U(1) easy-plane spin symmetry per flavor, resulting in a global $\mathrm{U}(\Nf) \times \mathrm{U}(\Nf)$ symmetry generated by the $2\Nf^2$ matrices ${\{\lambda_i\}_{i=1,\dots,\Nf^2} \otimes \{\mathbb{1}_4, \mathbb{1}_2 \otimes \sigma_z\}}$.

The construction of a Fierz-complete basis of four-fermion operators in the Gross-Neveu-Spin-XY theory space is fully analogous to the Gross-Neveu-Heisenberg case~\cite{Ladovrechis:2023}. A complete basis of four-fermion terms for the Gross-Neveu-Spin-XY model can be obtained from the complete basis of the Gross-Neveu-Heisenberg model by replacing all bilinears transforming as $\mathrm{SU}(2)$ vectors as
\begin{align}
    \frac{g}{2\Nf} [\bar{\psi} (\mathcal{N} \otimes \vec{\sigma}) \psi]^2 
& \mapsto 
    \frac{g_{xy}}{2\Nf} [\bar{\psi} (\mathcal{N} \otimes \vec{\sigma}_{xy}) \psi]^2 
\nonumber \\ & \quad
    + \frac{g_{z}}{2\Nf} [\bar{\psi} (\mathcal{N} \otimes{\sigma}_{z}) \psi]^2,
\end{align}
where $\vec \sigma_{xy} = (\sigma_x, \sigma_y)$ and $\mathcal N$ corresponds to a generic $2 \times 2$ operator.
As a result, a complete basis of four-fermion operators in the Gross-Neveu-Spin-XY theory space is given by 
\begin{align}\label{eq:Lintspin}
    \mathcal{L}_\mathrm{int}^\mathrm{spin} = &\frac{g_1}{2\Nf} [\bar{\psi} (\mathbb{1}_2 \otimes \mathbb{1}_2) \psi]^2 + \frac{g_2}{2\Nf} [\bar{\psi} (\Gamma_\mu \otimes \mathbb{1}_2) \psi]^2 \nonumber \\ 
    &+ \frac{g_3}{2\Nf} [\bar{\psi} (\Gamma_5 \otimes \sigma_{z}) \psi]^2 + \frac{g_4}{2\Nf} [\bar{\psi} (\Gamma_5 \otimes \vec{\sigma}_{xy}) \psi]^2 \nonumber\\
    &+ \frac{g_5}{2\Nf} [\bar{\psi} (\Gamma_5 \otimes \mathbb{1}_2) \psi]^2 + \frac{g_6}{2\Nf} [\bar{\psi} (\Gamma_\mu \otimes \sigma_{z}) \psi]^2 \nonumber\\
    &+ \frac{g_7}{2\Nf} [\bar{\psi} (\Gamma_\mu \otimes \vec{\sigma}_{xy}) \psi]^2 + \frac{g_8}{2\Nf} [\bar{\psi} (\mathbb{1}_2 \otimes \vec{\sigma}_{xy}) \psi]^2 \nonumber \\
    & + \frac{g_9}{2\Nf} [\bar{\psi} (\mathbb{1}_2 \otimes \sigma_{z}) \psi]^2,
\end{align}
where $\Gamma_5 = \rmi \Gamma_0 \Gamma_1$. 
Identifying $\gamma_\mu \coloneqq \Gamma_\mu \otimes \mathbb{1}_2$, $\gamma_2 \coloneqq \Gamma_5 \otimes \sigma_z$, $\gamma_3 \coloneqq \Gamma_5 \otimes \sigma_x$, and $\gamma_5 \coloneqq \Gamma_5 \otimes \sigma_y$ defines a reducible four-dimensional representation of the Clifford algebra.
A straightforward calculation reveals that this identification maps Eq.~\eqref{eq:Lintspin} to Eq.~\eqref{eq:Lintchi}. The Gross-Neveu-XY and Gross-Neveu-Spin-XY theory spaces are therefore equivalent.
Under the equivalence, the U(1) easy-plane spin symmetry is mapped to the continuous chiral symmetry generated by $\gamma_{35}$.
Importantly, the critical fixed point associated with the Gross-Neveu-Spin-XY model parametrized by $g_8$ maps to the GNXY$'$ fixed point of Eq.~\eqref{eq:GNXYprime-FP}. At one-loop order, the latter shares the same critical exponents with the Gross-Neveu-XY fixed point. Whether this property also hold at higher loop orders represents an interesting direction for future work.

%%%%%%%%%%%%%%%%%%%%%%%%%%%%%%%%%%%%%%%%%%%%%%%%%%%%%%%%%%%%%%%%%%%%%%%%%%%%%%%%
\section{Additional data}
\label{app:additional-data}
%%%%%%%%%%%%%%%%%%%%%%%%%%%%%%%%%%%%%%%%%%%%%%%%%%%%%%%%%%%%%%%%%%%%%%%%%%%%%%%%

In this appendix, we present additional results for the critical exponents of the Gross-Neveu-XY fixed point in $D=2+1$ space-time dimensions.
Table~\ref{tab:Nf8crits} shows the exponents for $\Nf = 8$ four-component Dirac fermions from different Pad\'e interpolations $[m/n]$ between the expansions near the lower critical dimension (Sec.~\ref{sec:chiralXY}) and the upper critical dimension (Ref.~\cite{Zerf4L:2017}).
Tables~\ref{tab:Nf4Pades} and \ref{tab:Nf2Pades} present the different one-sided Pad\'e approximants of the $4-\varepsilon$ results~\cite{Zerf4L:2017}, which were used to compute the averages shown in the second rows of Tables~\ref{tab:summaryNf=4} and \ref{tab:summaryNf=2}, respectively. Note that for $\Nf = 2$, the $[4/0]$ Pad\'e approximant estimates a negative value for $\omega$, suggesting that the fixed point is unstable, which would be consistent with our findings from the $2+\epsilon$ expansion. We emphasize, however, that the mechanism leading to the destabilization of the Gross-Neveu-XY fixed point as a function of $\Nf$ in the $4-\varepsilon$ expansion differs from the one discussed above within the $2+\epsilon$ expansion. The latter necessitates computing the scaling dimensions of four-fermion operators in condensation channels distinct from those analyzed in Ref.~\cite{Zerf4L:2017}.\\

%%%%%%%%%%%%%%%%%%%%%%%%%%%%%%%%%%%%%%%%%%%%%%%%%%%%%%%%%%%%%%%%%%%%%%%%%%%%%%%%
\begin{table}[tb!]
\caption{Same as Table~\ref{tab:Nf4crits}, but for $\Nf = 8$ four-component Dirac fermions.}
\begin{tabular*}{\linewidth}{@{\extracolsep{\fill}} l c l l l}
\noalign{\smallskip}\hline\hline\noalign{\smallskip}
 $\Nf = 8$ & $[m/n]$ & \multicolumn{1}{c}{$1/\nu$} & \multicolumn{1}{c}{$\eta_\varphi$} & \multicolumn{1}{c}{$\eta_\psi$} \\
\noalign{\smallskip}\hline\noalign{\smallskip}
 $\mathcal{O}(\epsilon,\varepsilon^4)$ & [1/5] & 0.950759 & 0.970499 & \multicolumn{1}{c}{n.e.}\\
 & [2/4] & 0.944568 & \multicolumn{1}{c}{sing.} &  \multicolumn{1}{c}{n.e.}\\
 & [3/3] & 0.939039 & \multicolumn{1}{c}{sing.} & 0.0191471 \\
 & [4/2] & 0.940347 & \multicolumn{1}{c}{sing.} & 0.0183759 \\
 & [5/1] & 0.944877 & 0.964375 & \multicolumn{1}{c}{sing.} \\
 & [6/0] & 0.953274 & 0.969479 & 0.018169 \\
 & \textbf{avg.} & \textbf{0.945(5)} & \textbf{0.968(3)} & \textbf{0.0185(5)}\\
\noalign{\smallskip}
 $\mathcal{O}(\epsilon^2,\varepsilon^4)$ & [3/4] & \multicolumn{1}{c}{--} & \multicolumn{1}{c}{--} & 0.0183647 \\
  & [4/3] & \multicolumn{1}{c}{--} & \multicolumn{1}{c}{--} & 0.0180923 \\
  & [5/2] & \multicolumn{1}{c}{--} & \multicolumn{1}{c}{--} & 0.018175 \\
  & [6/1] & \multicolumn{1}{c}{--} & \multicolumn{1}{c}{--} & 0.0183313 \\
  & [7/0] & \multicolumn{1}{c}{--} & \multicolumn{1}{c}{--} & 0.0183106 \\ 
  & \textbf{avg.} & \multicolumn{1}{c}{--} & \multicolumn{1}{c}{--} & \textbf{0.0182(1)} \\
\noalign{\smallskip}\hline\hline
\end{tabular*}
\label{tab:Nf8crits}
\end{table}
%%%%%%%%%%%%%%%%%%%%%%%%%%%%%%%%%%%%%%%%%%%%%%%%%%%%%%%%%%%%%%%%%%%%%%%%%%%%%%%%

%%%%%%%%%%%%%%%%%%%%%%%%%%%%%%%%%%%%%%%%%%%%%%%%%%%%%%%%%%%%%%%%%%%%%%%%%%%%%%%%
\begin{table}[tb!]
\caption{Critical exponents of the Gross-Neveu-XY fixed point in $D=2+1$ space-time dimensions for $\Nf = 4$ four-component Dirac fermions from one-sided Pad\'e approximants $[m/n]$ of the fourth-order expansion near the upper critical dimension~\cite{Zerf4L:2017}.}
\begin{tabular*}{\linewidth}{@{\extracolsep{\fill}} l c l l l l}
\noalign{\smallskip}\hline\hline\noalign{\smallskip}
 $\Nf = 4$ & $[m/n]$ & \multicolumn{1}{c}{$1/\nu$} & \multicolumn{1}{c}{$\eta_\varphi$} & \multicolumn{1}{c}{$\eta_\psi$} & \multicolumn{1}{c}{$\omega$} \\
\noalign{\smallskip}\hline\noalign{\smallskip}
 $\mathcal{O}(\varepsilon^4)$ 
 & [1/3] & 0.942997 & 1.030910 & 0.0492433 & 0.667957\\
 & [2/2] & 0.884678 & 0.929495 & 0.0465988 & 0.863714\\
 & [3/1] & 0.885092 & \multicolumn{1}{c}{sing.} & 0.0372515 & 0.861121\\
 & [4/0] & 1.045550 & 0.995211 & 0.0401420 & 0.396606\\
 & \textbf{avg.} & \textbf{0.94(8)} & \textbf{0.99(5)} & \textbf{0.043(5)} & \textbf{0.69(22)} \\
\noalign{\smallskip}\hline\hline
\end{tabular*}
\label{tab:Nf4Pades}
\end{table}
%%%%%%%%%%%%%%%%%%%%%%%%%%%%%%%%%%%%%%%%%%%%%%%%%%%%%%%%%%%%%%%%%%%%%%%%%%%%%%%%

%%%%%%%%%%%%%%%%%%%%%%%%%%%%%%%%%%%%%%%%%%%%%%%%%%%%%%%%%%%%%%%%%%%%%%%%%%%%%%%%
\begin{table}[tb!]
\caption{Same as Table~\ref{tab:Nf4Pades}, but for $\Nf = 2$ four-component Dirac fermions.}
\begin{tabular*}{\linewidth}{@{\extracolsep{\fill}} l c l l l l}
\noalign{\smallskip}\hline\hline\noalign{\smallskip}
 $\Nf = 2$ & $[m/n]$ & \multicolumn{1}{c}{$1/\nu$} & \multicolumn{1}{c}{$\eta_\varphi$} & \multicolumn{1}{c}{$\eta_\psi$} & \multicolumn{1}{c}{$\omega$} \\
\noalign{\smallskip}\hline\noalign{\smallskip}
 $\mathcal{O}(\varepsilon^4)$ 
 & [1/3] & 0.933401 & 1.191620 & 0.120718 & \phantom{$-$}0.429812\\
 & [2/2] & 0.839458 & 0.809872 & 0.117375 & \phantom{$-$}0.796114\\
 & [3/1] & 0.840525 & 0.787649 & 0.108208 & \phantom{$-$}0.780257\\
 & [4/0] & 1.147807 & 0.976529 & 0.126391 & $-0.603811$\\
 & \textbf{avg.} & \textbf{0.94(15)} & \textbf{0.94(19)} & \textbf{0.118(8)} & \phantom{$-$}\textbf{0.35(66)}\\
\noalign{\smallskip}\hline\hline
\end{tabular*}
\label{tab:Nf2Pades}
\end{table}
%%%%%%%%%%%%%%%%%%%%%%%%%%%%%%%%%%%%%%%%%%%%%%%%%%%%%%%%%%%%%%%%%%%%%%%%%%%%%%%%

\FloatBarrier
\bibliographystyle{longapsrev4-2}
\bibliography{gn-xy}

%longapsrev4-2.bst hand-edited version by Lukas Janssen of apsrev4-2.bst
\begin{thebibliography}{58}%
\makeatletter
\providecommand \@ifxundefined [1]{%
 \@ifx{#1\undefined}
}%
\providecommand \@ifnum [1]{%
 \ifnum #1\expandafter \@firstoftwo
 \else \expandafter \@secondoftwo
 \fi
}%
\providecommand \@ifx [1]{%
 \ifx #1\expandafter \@firstoftwo
 \else \expandafter \@secondoftwo
 \fi
}%
\providecommand \natexlab [1]{#1}%
\providecommand \enquote  [1]{``#1''}%
\providecommand \bibnamefont  [1]{#1}%
\providecommand \bibfnamefont [1]{#1}%
\providecommand \citenamefont [1]{#1}%
\providecommand \href@noop [0]{\@secondoftwo}%
\providecommand \href [0]{\begingroup \@sanitize@url \@href}%
\providecommand \@href[1]{\@@startlink{#1}\@@href}%
\providecommand \@@href[1]{\endgroup#1\@@endlink}%
\providecommand \@sanitize@url [0]{\catcode `\\12\catcode `\$12\catcode
  `\&12\catcode `\#12\catcode `\^12\catcode `\_12\catcode `\%12\relax}%
\providecommand \@@startlink[1]{}%
\providecommand \@@endlink[0]{}%
\providecommand \url  [0]{\begingroup\@sanitize@url \@url }%
\providecommand \@url [1]{\endgroup\@href {#1}{\urlprefix }}%
\providecommand \urlprefix  [0]{URL }%
\providecommand \Eprint [0]{\href }%
\providecommand \doibase [0]{https://doi.org/}%
\providecommand \selectlanguage [0]{\@gobble}%
\providecommand \bibinfo  [0]{\@secondoftwo}%
\providecommand \bibfield  [0]{\@secondoftwo}%
\providecommand \translation [1]{[#1]}%
\providecommand \BibitemOpen [0]{}%
\providecommand \bibitemStop [0]{}%
\providecommand \bibitemNoStop [0]{.\EOS\space}%
\providecommand \EOS [0]{\spacefactor3000\relax}%
\providecommand \BibitemShut  [1]{\csname bibitem#1\endcsname}%
\let\auto@bib@innerbib\@empty
%</preamble>
\bibitem [{\citenamefont {Katsnelson}(2007)}]{Katsnelson:2007}%
  \BibitemOpen
  \bibfield  {author} {\bibinfo {author} {\bibfnamefont {M.~I.}\ \bibnamefont
  {Katsnelson}},\ }\bibfield  {title} {\bibinfo {title} {Graphene: carbon in
  two dimensions},\ }\href {https://doi.org/10.1016/s1369-7021(06)71788-6}
  {\bibfield  {journal} {\bibinfo  {journal} {Mater. Today}\ }\textbf {\bibinfo
  {volume} {10}},\ \bibinfo {pages} {20–27} (\bibinfo {year}
  {2007})}\BibitemShut {NoStop}%
\bibitem [{\citenamefont {Sorella}\ and\ \citenamefont
  {Tosatti}(1992)}]{Sorella:1992}%
  \BibitemOpen
  \bibfield  {author} {\bibinfo {author} {\bibfnamefont {S.}~\bibnamefont
  {Sorella}}\ and\ \bibinfo {author} {\bibfnamefont {E.}~\bibnamefont
  {Tosatti}},\ }\bibfield  {title} {\bibinfo {title} {Semi-Metal-Insulator
  Transition of the Hubbard Model in the Honeycomb Lattice},\ }\href
  {https://doi.org/10.1209/0295-5075/19/8/007} {\bibfield  {journal} {\bibinfo
  {journal} {Europhys. Lett.}\ }\textbf {\bibinfo {volume} {19}},\ \bibinfo
  {pages} {699} (\bibinfo {year} {1992})}\BibitemShut {NoStop}%
\bibitem [{\citenamefont {Martelo}\ \emph {et~al.}(1996)\citenamefont
  {Martelo}, \citenamefont {Dzierzawa}, \citenamefont {Siffert},\ and\
  \citenamefont {Baeriswyl}}]{Martelo:1996}%
  \BibitemOpen
  \bibfield  {author} {\bibinfo {author} {\bibfnamefont {L.~M.}\ \bibnamefont
  {Martelo}}, \bibinfo {author} {\bibfnamefont {M.}~\bibnamefont {Dzierzawa}},
  \bibinfo {author} {\bibfnamefont {L.}~\bibnamefont {Siffert}},\ and\ \bibinfo
  {author} {\bibfnamefont {D.}~\bibnamefont {Baeriswyl}},\ }\bibfield  {title}
  {\bibinfo {title} {Mott-Hubbard transition and antiferromagnetism on the
  honeycomb lattice},\ }\href {https://doi.org/10.1007/s002570050384}
  {\bibfield  {journal} {\bibinfo  {journal} {Z. Phys. B Condens. Matter}\
  }\textbf {\bibinfo {volume} {103}},\ \bibinfo {pages} {335} (\bibinfo {year}
  {1996})}\BibitemShut {NoStop}%
\bibitem [{\citenamefont {Paiva}\ \emph {et~al.}(2005)\citenamefont {Paiva},
  \citenamefont {Scalettar}, \citenamefont {Zheng}, \citenamefont {Singh},\
  and\ \citenamefont {Oitmaa}}]{Paiva:2005}%
  \BibitemOpen
  \bibfield  {author} {\bibinfo {author} {\bibfnamefont {T.}~\bibnamefont
  {Paiva}}, \bibinfo {author} {\bibfnamefont {R.~T.}\ \bibnamefont
  {Scalettar}}, \bibinfo {author} {\bibfnamefont {W.}~\bibnamefont {Zheng}},
  \bibinfo {author} {\bibfnamefont {R.~R.~P.}\ \bibnamefont {Singh}},\ and\
  \bibinfo {author} {\bibfnamefont {J.}~\bibnamefont {Oitmaa}},\ }\bibfield
  {title} {\bibinfo {title} {Ground-state and finite-temperature signatures of
  quantum phase transitions in the half-filled Hubbard model on a honeycomb
  lattice},\ }\href {https://doi.org/10.1103/PhysRevB.72.085123} {\bibfield
  {journal} {\bibinfo  {journal} {Phys. Rev. B}\ }\textbf {\bibinfo {volume}
  {72}},\ \bibinfo {pages} {085123} (\bibinfo {year} {2005})}\BibitemShut
  {NoStop}%
\bibitem [{\citenamefont {Herbut}(2006)}]{HerbutGraphene:2006}%
  \BibitemOpen
  \bibfield  {author} {\bibinfo {author} {\bibfnamefont {I.~F.}\ \bibnamefont
  {Herbut}},\ }\bibfield  {title} {\bibinfo {title} {Interactions and Phase
  Transitions on Graphene's Honeycomb Lattice},\ }\href
  {https://doi.org/10.1103/PhysRevLett.97.146401} {\bibfield  {journal}
  {\bibinfo  {journal} {Phys. Rev. Lett.}\ }\textbf {\bibinfo {volume} {97}},\
  \bibinfo {pages} {146401} (\bibinfo {year} {2006})}\BibitemShut {NoStop}%
\bibitem [{\citenamefont {Assaad}\ and\ \citenamefont
  {Herbut}(2013)}]{assaad13}%
  \BibitemOpen
  \bibfield  {author} {\bibinfo {author} {\bibfnamefont {F.~F.}\ \bibnamefont
  {Assaad}}\ and\ \bibinfo {author} {\bibfnamefont {I.~F.}\ \bibnamefont
  {Herbut}},\ }\bibfield  {title} {\bibinfo {title} {Pinning the Order: The
  Nature of Quantum Criticality in the Hubbard Model on Honeycomb Lattice},\
  }\href {https://doi.org/10.1103/PhysRevX.3.031010} {\bibfield  {journal}
  {\bibinfo  {journal} {Phys. Rev. X}\ }\textbf {\bibinfo {volume} {3}},\
  \bibinfo {pages} {031010} (\bibinfo {year} {2013})}\BibitemShut {NoStop}%
\bibitem [{\citenamefont {Raghu}\ \emph {et~al.}(2008)\citenamefont {Raghu},
  \citenamefont {Qi}, \citenamefont {Honerkamp},\ and\ \citenamefont
  {Zhang}}]{raghu08}%
  \BibitemOpen
  \bibfield  {author} {\bibinfo {author} {\bibfnamefont {S.}~\bibnamefont
  {Raghu}}, \bibinfo {author} {\bibfnamefont {X.-L.}\ \bibnamefont {Qi}},
  \bibinfo {author} {\bibfnamefont {C.}~\bibnamefont {Honerkamp}},\ and\
  \bibinfo {author} {\bibfnamefont {S.-C.}\ \bibnamefont {Zhang}},\ }\bibfield
  {title} {\bibinfo {title} {Topological Mott Insulators},\ }\href
  {https://doi.org/10.1103/PhysRevLett.100.156401} {\bibfield  {journal}
  {\bibinfo  {journal} {Phys. Rev. Lett.}\ }\textbf {\bibinfo {volume} {100}},\
  \bibinfo {pages} {156401} (\bibinfo {year} {2008})}\BibitemShut {NoStop}%
\bibitem [{\citenamefont {Herbut}\ \emph {et~al.}(2009)\citenamefont {Herbut},
  \citenamefont {Juri\ifmmode \check{c}\else \v{c}\fi{}i\ifmmode~\acute{c}\else
  \'{c}\fi{}},\ and\ \citenamefont {Roy}}]{HerbutInteracting:2009}%
  \BibitemOpen
  \bibfield  {author} {\bibinfo {author} {\bibfnamefont {I.~F.}\ \bibnamefont
  {Herbut}}, \bibinfo {author} {\bibfnamefont {V.}~\bibnamefont {Juri\ifmmode
  \check{c}\else \v{c}\fi{}i\ifmmode~\acute{c}\else \'{c}\fi{}}},\ and\
  \bibinfo {author} {\bibfnamefont {B.}~\bibnamefont {Roy}},\ }\bibfield
  {title} {\bibinfo {title} {Theory of interacting electrons on the honeycomb
  lattice},\ }\href {https://doi.org/10.1103/PhysRevB.79.085116} {\bibfield
  {journal} {\bibinfo  {journal} {Phys. Rev. B}\ }\textbf {\bibinfo {volume}
  {79}},\ \bibinfo {pages} {085116} (\bibinfo {year} {2009})}\BibitemShut
  {NoStop}%
\bibitem [{\citenamefont {Weeks}\ and\ \citenamefont {Franz}(2010)}]{weeks10}%
  \BibitemOpen
  \bibfield  {author} {\bibinfo {author} {\bibfnamefont {C.}~\bibnamefont
  {Weeks}}\ and\ \bibinfo {author} {\bibfnamefont {M.}~\bibnamefont {Franz}},\
  }\bibfield  {title} {\bibinfo {title} {Interaction-driven instabilities of a
  Dirac semimetal},\ }\href {https://doi.org/10.1103/PhysRevB.81.085105}
  {\bibfield  {journal} {\bibinfo  {journal} {Phys. Rev. B}\ }\textbf {\bibinfo
  {volume} {81}},\ \bibinfo {pages} {085105} (\bibinfo {year}
  {2010})}\BibitemShut {NoStop}%
\bibitem [{\citenamefont {Scherer}\ and\ \citenamefont
  {Herbut}(2016)}]{Scherer:2016}%
  \BibitemOpen
  \bibfield  {author} {\bibinfo {author} {\bibfnamefont {M.~M.}\ \bibnamefont
  {Scherer}}\ and\ \bibinfo {author} {\bibfnamefont {I.~F.}\ \bibnamefont
  {Herbut}},\ }\bibfield  {title} {\bibinfo {title} {Gauge-field-assisted
  Kekul\'e quantum criticality},\ }\href
  {https://doi.org/10.1103/PhysRevB.94.205136} {\bibfield  {journal} {\bibinfo
  {journal} {Phys. Rev. B}\ }\textbf {\bibinfo {volume} {94}},\ \bibinfo
  {pages} {205136} (\bibinfo {year} {2016})}\BibitemShut {NoStop}%
\bibitem [{\citenamefont {Li}\ \emph {et~al.}(2017)\citenamefont {Li},
  \citenamefont {Jiang}, \citenamefont {Jian},\ and\ \citenamefont
  {Yao}}]{LiFermi:2017}%
  \BibitemOpen
  \bibfield  {author} {\bibinfo {author} {\bibfnamefont {Z.-X.}\ \bibnamefont
  {Li}}, \bibinfo {author} {\bibfnamefont {Y.-F.}\ \bibnamefont {Jiang}},
  \bibinfo {author} {\bibfnamefont {S.-K.}\ \bibnamefont {Jian}},\ and\
  \bibinfo {author} {\bibfnamefont {H.}~\bibnamefont {Yao}},\ }\bibfield
  {title} {\bibinfo {title} {Fermion-induced quantum critical points},\ }\href
  {https://doi.org/10.1038/s41467-017-00167-6} {\bibfield  {journal} {\bibinfo
  {journal} {Nat. Commun.}\ }\textbf {\bibinfo {volume} {8}},\ \bibinfo {pages}
  {314} (\bibinfo {year} {2017})}\BibitemShut {NoStop}%
\bibitem [{\citenamefont {Xu}\ \emph {et~al.}(2018)\citenamefont {Xu},
  \citenamefont {Law},\ and\ \citenamefont {Lee}}]{Xu:2018}%
  \BibitemOpen
  \bibfield  {author} {\bibinfo {author} {\bibfnamefont {X.~Y.}\ \bibnamefont
  {Xu}}, \bibinfo {author} {\bibfnamefont {K.~T.}\ \bibnamefont {Law}},\ and\
  \bibinfo {author} {\bibfnamefont {P.~A.}\ \bibnamefont {Lee}},\ }\bibfield
  {title} {\bibinfo {title} {Kekul\'e valence bond order in an extended Hubbard
  model on the honeycomb lattice with possible applications to twisted bilayer
  graphene},\ }\href {https://doi.org/10.1103/PhysRevB.98.121406} {\bibfield
  {journal} {\bibinfo  {journal} {Phys. Rev. B}\ }\textbf {\bibinfo {volume}
  {98}},\ \bibinfo {pages} {121406} (\bibinfo {year} {2018})}\BibitemShut
  {NoStop}%
\bibitem [{\citenamefont {Roy}\ and\ \citenamefont
  {Herbut}(2010)}]{RoyKekule:2010}%
  \BibitemOpen
  \bibfield  {author} {\bibinfo {author} {\bibfnamefont {B.}~\bibnamefont
  {Roy}}\ and\ \bibinfo {author} {\bibfnamefont {I.~F.}\ \bibnamefont
  {Herbut}},\ }\bibfield  {title} {\bibinfo {title} {Unconventional
  superconductivity on honeycomb lattice: Theory of Kekule order parameter},\
  }\href {https://doi.org/10.1103/PhysRevB.82.035429} {\bibfield  {journal}
  {\bibinfo  {journal} {Phys. Rev. B}\ }\textbf {\bibinfo {volume} {82}},\
  \bibinfo {pages} {035429} (\bibinfo {year} {2010})}\BibitemShut {NoStop}%
\bibitem [{\citenamefont {Otsuka}\ \emph {et~al.}(2018)\citenamefont {Otsuka},
  \citenamefont {Seki}, \citenamefont {Sorella},\ and\ \citenamefont
  {Yunoki}}]{Otsuka:2018}%
  \BibitemOpen
  \bibfield  {author} {\bibinfo {author} {\bibfnamefont {Y.}~\bibnamefont
  {Otsuka}}, \bibinfo {author} {\bibfnamefont {K.}~\bibnamefont {Seki}},
  \bibinfo {author} {\bibfnamefont {S.}~\bibnamefont {Sorella}},\ and\ \bibinfo
  {author} {\bibfnamefont {S.}~\bibnamefont {Yunoki}},\ }\bibfield  {title}
  {\bibinfo {title} {Quantum criticality in the metal-superconductor transition
  of interacting Dirac fermions on a triangular lattice},\ }\href
  {https://doi.org/10.1103/PhysRevB.98.035126} {\bibfield  {journal} {\bibinfo
  {journal} {Phys. Rev. B}\ }\textbf {\bibinfo {volume} {98}},\ \bibinfo
  {pages} {035126} (\bibinfo {year} {2018})}\BibitemShut {NoStop}%
\bibitem [{\citenamefont {Gehring}\ \emph {et~al.}(2015)\citenamefont
  {Gehring}, \citenamefont {Gies},\ and\ \citenamefont
  {Janssen}}]{Gehring:2015}%
  \BibitemOpen
  \bibfield  {author} {\bibinfo {author} {\bibfnamefont {F.}~\bibnamefont
  {Gehring}}, \bibinfo {author} {\bibfnamefont {H.}~\bibnamefont {Gies}},\ and\
  \bibinfo {author} {\bibfnamefont {L.}~\bibnamefont {Janssen}},\ }\bibfield
  {title} {\bibinfo {title} {Fixed-point structure of low-dimensional
  relativistic fermion field theories: Universality classes and emergent
  symmetry},\ }\href {https://doi.org/10.1103/PhysRevD.92.085046} {\bibfield
  {journal} {\bibinfo  {journal} {Phys. Rev. D}\ }\textbf {\bibinfo {volume}
  {92}},\ \bibinfo {pages} {085046} (\bibinfo {year} {2015})}\BibitemShut
  {NoStop}%
\bibitem [{\citenamefont {Zerf}\ \emph {et~al.}(2017)\citenamefont {Zerf},
  \citenamefont {Mihaila}, \citenamefont {Marquard}, \citenamefont {Herbut},\
  and\ \citenamefont {Scherer}}]{Zerf4L:2017}%
  \BibitemOpen
  \bibfield  {author} {\bibinfo {author} {\bibfnamefont {N.}~\bibnamefont
  {Zerf}}, \bibinfo {author} {\bibfnamefont {L.~N.}\ \bibnamefont {Mihaila}},
  \bibinfo {author} {\bibfnamefont {P.}~\bibnamefont {Marquard}}, \bibinfo
  {author} {\bibfnamefont {I.~F.}\ \bibnamefont {Herbut}},\ and\ \bibinfo
  {author} {\bibfnamefont {M.~M.}\ \bibnamefont {Scherer}},\ }\bibfield
  {title} {\bibinfo {title} {Four-loop critical exponents for the
  Gross-Neveu-Yukawa models},\ }\href
  {https://doi.org/10.1103/PhysRevD.96.096010} {\bibfield  {journal} {\bibinfo
  {journal} {Phys. Rev. D}\ }\textbf {\bibinfo {volume} {96}},\ \bibinfo
  {pages} {096010} (\bibinfo {year} {2017})}\BibitemShut {NoStop}%
\bibitem [{\citenamefont {Herbut}(2024)}]{herbut24}%
  \BibitemOpen
  \bibfield  {author} {\bibinfo {author} {\bibfnamefont {I.~F.}\ \bibnamefont
  {Herbut}},\ }\bibfield  {title} {\bibinfo {title} {Wilson-Fisher fixed points
  in the presence of Dirac fermions},\ }\href
  {https://doi.org/10.1142/S0217984924300060} {\bibfield  {journal} {\bibinfo
  {journal} {Mod. Phys. Lett. B}\ }\textbf {\bibinfo {volume} {38}},\ \bibinfo
  {pages} {2430006} (\bibinfo {year} {2024})}\BibitemShut {NoStop}%
\bibitem [{\citenamefont {Shallcross}\ \emph {et~al.}(2010)\citenamefont
  {Shallcross}, \citenamefont {Sharma}, \citenamefont {Kandelaki},\ and\
  \citenamefont {Pankratov}}]{Shallcross:2010}%
  \BibitemOpen
  \bibfield  {author} {\bibinfo {author} {\bibfnamefont {S.}~\bibnamefont
  {Shallcross}}, \bibinfo {author} {\bibfnamefont {S.}~\bibnamefont {Sharma}},
  \bibinfo {author} {\bibfnamefont {E.}~\bibnamefont {Kandelaki}},\ and\
  \bibinfo {author} {\bibfnamefont {O.~A.}\ \bibnamefont {Pankratov}},\
  }\bibfield  {title} {\bibinfo {title} {Electronic structure of turbostratic
  graphene},\ }\href {https://doi.org/10.1103/PhysRevB.81.165105} {\bibfield
  {journal} {\bibinfo  {journal} {Phys. Rev. B}\ }\textbf {\bibinfo {volume}
  {81}},\ \bibinfo {pages} {165105} (\bibinfo {year} {2010})}\BibitemShut
  {NoStop}%
\bibitem [{\citenamefont {Bistritzer}\ and\ \citenamefont
  {MacDonald}(2011)}]{Bistritzer:2011}%
  \BibitemOpen
  \bibfield  {author} {\bibinfo {author} {\bibfnamefont {R.}~\bibnamefont
  {Bistritzer}}\ and\ \bibinfo {author} {\bibfnamefont {A.~H.}\ \bibnamefont
  {MacDonald}},\ }\bibfield  {title} {\bibinfo {title} {Moir{\'e} bands in
  twisted double-layer graphene},\ }\href
  {https://doi.org/10.1073/pnas.1108174108} {\bibfield  {journal} {\bibinfo
  {journal} {Proc. Natl. Acad. Sci.}\ }\textbf {\bibinfo {volume} {108}},\
  \bibinfo {pages} {12233} (\bibinfo {year} {2011})}\BibitemShut {NoStop}%
\bibitem [{\citenamefont {Luican}\ \emph {et~al.}(2011)\citenamefont {Luican},
  \citenamefont {Li}, \citenamefont {Reina}, \citenamefont {Kong},
  \citenamefont {Nair}, \citenamefont {Novoselov}, \citenamefont {Geim},\ and\
  \citenamefont {Andrei}}]{Luican:2011}%
  \BibitemOpen
  \bibfield  {author} {\bibinfo {author} {\bibfnamefont {A.}~\bibnamefont
  {Luican}}, \bibinfo {author} {\bibfnamefont {G.}~\bibnamefont {Li}}, \bibinfo
  {author} {\bibfnamefont {A.}~\bibnamefont {Reina}}, \bibinfo {author}
  {\bibfnamefont {J.}~\bibnamefont {Kong}}, \bibinfo {author} {\bibfnamefont
  {R.~R.}\ \bibnamefont {Nair}}, \bibinfo {author} {\bibfnamefont {K.~S.}\
  \bibnamefont {Novoselov}}, \bibinfo {author} {\bibfnamefont {A.~K.}\
  \bibnamefont {Geim}},\ and\ \bibinfo {author} {\bibfnamefont {E.~Y.}\
  \bibnamefont {Andrei}},\ }\bibfield  {title} {\bibinfo {title} {Single-Layer
  Behavior and Its Breakdown in Twisted Graphene Layers},\ }\href
  {https://doi.org/10.1103/PhysRevLett.106.126802} {\bibfield  {journal}
  {\bibinfo  {journal} {Phys. Rev. Lett.}\ }\textbf {\bibinfo {volume} {106}},\
  \bibinfo {pages} {126802} (\bibinfo {year} {2011})}\BibitemShut {NoStop}%
\bibitem [{\citenamefont {Sanchez-Yamagishi}\ \emph {et~al.}(2012)\citenamefont
  {Sanchez-Yamagishi}, \citenamefont {Taychatanapat}, \citenamefont {Watanabe},
  \citenamefont {Taniguchi}, \citenamefont {Yacoby},\ and\ \citenamefont
  {Jarillo-Herrero}}]{Sanchez:2012}%
  \BibitemOpen
  \bibfield  {author} {\bibinfo {author} {\bibfnamefont {J.~D.}\ \bibnamefont
  {Sanchez-Yamagishi}}, \bibinfo {author} {\bibfnamefont {T.}~\bibnamefont
  {Taychatanapat}}, \bibinfo {author} {\bibfnamefont {K.}~\bibnamefont
  {Watanabe}}, \bibinfo {author} {\bibfnamefont {T.}~\bibnamefont {Taniguchi}},
  \bibinfo {author} {\bibfnamefont {A.}~\bibnamefont {Yacoby}},\ and\ \bibinfo
  {author} {\bibfnamefont {P.}~\bibnamefont {Jarillo-Herrero}},\ }\bibfield
  {title} {\bibinfo {title} {Quantum Hall Effect, Screening, and
  Layer-Polarized Insulating States in Twisted Bilayer Graphene},\ }\href
  {https://doi.org/10.1103/PhysRevLett.108.076601} {\bibfield  {journal}
  {\bibinfo  {journal} {Phys. Rev. Lett.}\ }\textbf {\bibinfo {volume} {108}},\
  \bibinfo {pages} {076601} (\bibinfo {year} {2012})}\BibitemShut {NoStop}%
\bibitem [{\citenamefont {Cao}\ \emph {et~al.}(2016)\citenamefont {Cao},
  \citenamefont {Luo}, \citenamefont {Fatemi}, \citenamefont {Fang},
  \citenamefont {Sanchez-Yamagishi}, \citenamefont {Watanabe}, \citenamefont
  {Taniguchi}, \citenamefont {Kaxiras},\ and\ \citenamefont
  {Jarillo-Herrero}}]{Cao:2016}%
  \BibitemOpen
  \bibfield  {author} {\bibinfo {author} {\bibfnamefont {Y.}~\bibnamefont
  {Cao}}, \bibinfo {author} {\bibfnamefont {J.~Y.}\ \bibnamefont {Luo}},
  \bibinfo {author} {\bibfnamefont {V.}~\bibnamefont {Fatemi}}, \bibinfo
  {author} {\bibfnamefont {S.}~\bibnamefont {Fang}}, \bibinfo {author}
  {\bibfnamefont {J.~D.}\ \bibnamefont {Sanchez-Yamagishi}}, \bibinfo {author}
  {\bibfnamefont {K.}~\bibnamefont {Watanabe}}, \bibinfo {author}
  {\bibfnamefont {T.}~\bibnamefont {Taniguchi}}, \bibinfo {author}
  {\bibfnamefont {E.}~\bibnamefont {Kaxiras}},\ and\ \bibinfo {author}
  {\bibfnamefont {P.}~\bibnamefont {Jarillo-Herrero}},\ }\bibfield  {title}
  {\bibinfo {title} {Superlattice-Induced Insulating States and
  Valley-Protected Orbits in Twisted Bilayer Graphene},\ }\href
  {https://doi.org/10.1103/PhysRevLett.117.116804} {\bibfield  {journal}
  {\bibinfo  {journal} {Phys. Rev. Lett.}\ }\textbf {\bibinfo {volume} {117}},\
  \bibinfo {pages} {116804} (\bibinfo {year} {2016})}\BibitemShut {NoStop}%
\bibitem [{\citenamefont {Lopes~dos Santos}\ \emph {et~al.}(2012)\citenamefont
  {Lopes~dos Santos}, \citenamefont {Peres},\ and\ \citenamefont
  {Castro~Neto}}]{Lopes:2012}%
  \BibitemOpen
  \bibfield  {author} {\bibinfo {author} {\bibfnamefont {J.~M.~B.}\
  \bibnamefont {Lopes~dos Santos}}, \bibinfo {author} {\bibfnamefont
  {N.~M.~R.}\ \bibnamefont {Peres}},\ and\ \bibinfo {author} {\bibfnamefont
  {A.~H.}\ \bibnamefont {Castro~Neto}},\ }\bibfield  {title} {\bibinfo {title}
  {Continuum model of the twisted graphene bilayer},\ }\href
  {https://doi.org/10.1103/PhysRevB.86.155449} {\bibfield  {journal} {\bibinfo
  {journal} {Phys. Rev. B}\ }\textbf {\bibinfo {volume} {86}},\ \bibinfo
  {pages} {155449} (\bibinfo {year} {2012})}\BibitemShut {NoStop}%
\bibitem [{\citenamefont {Tarnopolsky}\ \emph {et~al.}(2019)\citenamefont
  {Tarnopolsky}, \citenamefont {Kruchkov},\ and\ \citenamefont
  {Vishwanath}}]{Tarnopolsky:2019}%
  \BibitemOpen
  \bibfield  {author} {\bibinfo {author} {\bibfnamefont {G.}~\bibnamefont
  {Tarnopolsky}}, \bibinfo {author} {\bibfnamefont {A.~J.}\ \bibnamefont
  {Kruchkov}},\ and\ \bibinfo {author} {\bibfnamefont {A.}~\bibnamefont
  {Vishwanath}},\ }\bibfield  {title} {\bibinfo {title} {Origin of Magic Angles
  in Twisted Bilayer Graphene},\ }\href
  {https://doi.org/10.1103/PhysRevLett.122.106405} {\bibfield  {journal}
  {\bibinfo  {journal} {Phys. Rev. Lett.}\ }\textbf {\bibinfo {volume} {122}},\
  \bibinfo {pages} {106405} (\bibinfo {year} {2019})}\BibitemShut {NoStop}%
\bibitem [{\citenamefont {Bernevig}\ \emph {et~al.}(2021)\citenamefont
  {Bernevig}, \citenamefont {Song}, \citenamefont {Regnault},\ and\
  \citenamefont {Lian}}]{Bernevig:2021}%
  \BibitemOpen
  \bibfield  {author} {\bibinfo {author} {\bibfnamefont {B.~A.}\ \bibnamefont
  {Bernevig}}, \bibinfo {author} {\bibfnamefont {Z.-D.}\ \bibnamefont {Song}},
  \bibinfo {author} {\bibfnamefont {N.}~\bibnamefont {Regnault}},\ and\
  \bibinfo {author} {\bibfnamefont {B.}~\bibnamefont {Lian}},\ }\bibfield
  {title} {\bibinfo {title} {Twisted bilayer graphene. III. Interacting
  Hamiltonian and exact symmetries},\ }\href
  {https://doi.org/10.1103/PhysRevB.103.205413} {\bibfield  {journal} {\bibinfo
   {journal} {Phys. Rev. B}\ }\textbf {\bibinfo {volume} {103}},\ \bibinfo
  {pages} {205413} (\bibinfo {year} {2021})}\BibitemShut {NoStop}%
\bibitem [{\citenamefont {Lu}\ \emph {et~al.}(2019)\citenamefont {Lu},
  \citenamefont {Stepanov}, \citenamefont {Yang}, \citenamefont {Xie},
  \citenamefont {Aamir}, \citenamefont {Das}, \citenamefont {Urgell},
  \citenamefont {Watanabe}, \citenamefont {Taniguchi}, \citenamefont {Zhang},
  \citenamefont {Bachtold}, \citenamefont {MacDonald},\ and\ \citenamefont
  {Efetov}}]{Lu:2019}%
  \BibitemOpen
  \bibfield  {author} {\bibinfo {author} {\bibfnamefont {X.}~\bibnamefont
  {Lu}}, \bibinfo {author} {\bibfnamefont {P.}~\bibnamefont {Stepanov}},
  \bibinfo {author} {\bibfnamefont {W.}~\bibnamefont {Yang}}, \bibinfo {author}
  {\bibfnamefont {M.}~\bibnamefont {Xie}}, \bibinfo {author} {\bibfnamefont
  {M.~A.}\ \bibnamefont {Aamir}}, \bibinfo {author} {\bibfnamefont
  {I.}~\bibnamefont {Das}}, \bibinfo {author} {\bibfnamefont {C.}~\bibnamefont
  {Urgell}}, \bibinfo {author} {\bibfnamefont {K.}~\bibnamefont {Watanabe}},
  \bibinfo {author} {\bibfnamefont {T.}~\bibnamefont {Taniguchi}}, \bibinfo
  {author} {\bibfnamefont {G.}~\bibnamefont {Zhang}}, \bibinfo {author}
  {\bibfnamefont {A.}~\bibnamefont {Bachtold}}, \bibinfo {author}
  {\bibfnamefont {A.~H.}\ \bibnamefont {MacDonald}},\ and\ \bibinfo {author}
  {\bibfnamefont {D.~K.}\ \bibnamefont {Efetov}},\ }\bibfield  {title}
  {\bibinfo {title} {Superconductors, orbital magnets and correlated states in
  magic-angle bilayer graphene},\ }\href
  {https://doi.org/10.1038/s41586-019-1695-0} {\bibfield  {journal} {\bibinfo
  {journal} {Nature}\ }\textbf {\bibinfo {volume} {574}},\ \bibinfo {pages}
  {653} (\bibinfo {year} {2019})}\BibitemShut {NoStop}%
\bibitem [{\citenamefont {Cao}\ \emph {et~al.}(2018)\citenamefont {Cao},
  \citenamefont {Fatemi}, \citenamefont {Demir}, \citenamefont {Fang},
  \citenamefont {Tomarken}, \citenamefont {Luo}, \citenamefont
  {Sanchez-Yamagishi}, \citenamefont {Watanabe}, \citenamefont {Taniguchi},
  \citenamefont {Kaxiras}, \citenamefont {Ashoori},\ and\ \citenamefont
  {Jarillo-Herrero}}]{Cao:2018}%
  \BibitemOpen
  \bibfield  {author} {\bibinfo {author} {\bibfnamefont {Y.}~\bibnamefont
  {Cao}}, \bibinfo {author} {\bibfnamefont {V.}~\bibnamefont {Fatemi}},
  \bibinfo {author} {\bibfnamefont {A.}~\bibnamefont {Demir}}, \bibinfo
  {author} {\bibfnamefont {S.}~\bibnamefont {Fang}}, \bibinfo {author}
  {\bibfnamefont {S.~L.}\ \bibnamefont {Tomarken}}, \bibinfo {author}
  {\bibfnamefont {J.~Y.}\ \bibnamefont {Luo}}, \bibinfo {author} {\bibfnamefont
  {J.~D.}\ \bibnamefont {Sanchez-Yamagishi}}, \bibinfo {author} {\bibfnamefont
  {K.}~\bibnamefont {Watanabe}}, \bibinfo {author} {\bibfnamefont
  {T.}~\bibnamefont {Taniguchi}}, \bibinfo {author} {\bibfnamefont
  {E.}~\bibnamefont {Kaxiras}}, \bibinfo {author} {\bibfnamefont {R.~C.}\
  \bibnamefont {Ashoori}},\ and\ \bibinfo {author} {\bibfnamefont
  {P.}~\bibnamefont {Jarillo-Herrero}},\ }\bibfield  {title} {\bibinfo {title}
  {Correlated insulator behaviour at half-filling in magic-angle graphene
  superlattices},\ }\href {https://doi.org/10.1038/nature26154} {\bibfield
  {journal} {\bibinfo  {journal} {Nature}\ }\textbf {\bibinfo {volume} {556}},\
  \bibinfo {pages} {80} (\bibinfo {year} {2018})}\BibitemShut {NoStop}%
\bibitem [{\citenamefont {Po}\ \emph {et~al.}(2018)\citenamefont {Po},
  \citenamefont {Zou}, \citenamefont {Vishwanath},\ and\ \citenamefont
  {Senthil}}]{Po:2018}%
  \BibitemOpen
  \bibfield  {author} {\bibinfo {author} {\bibfnamefont {H.~C.}\ \bibnamefont
  {Po}}, \bibinfo {author} {\bibfnamefont {L.}~\bibnamefont {Zou}}, \bibinfo
  {author} {\bibfnamefont {A.}~\bibnamefont {Vishwanath}},\ and\ \bibinfo
  {author} {\bibfnamefont {T.}~\bibnamefont {Senthil}},\ }\bibfield  {title}
  {\bibinfo {title} {Origin of Mott Insulating Behavior and Superconductivity
  in Twisted Bilayer Graphene},\ }\href
  {https://doi.org/10.1103/PhysRevX.8.031089} {\bibfield  {journal} {\bibinfo
  {journal} {Phys. Rev. X}\ }\textbf {\bibinfo {volume} {8}},\ \bibinfo {pages}
  {031089} (\bibinfo {year} {2018})}\BibitemShut {NoStop}%
\bibitem [{\citenamefont {Bultinck}\ \emph {et~al.}(2020)\citenamefont
  {Bultinck}, \citenamefont {Khalaf}, \citenamefont {Liu}, \citenamefont
  {Chatterjee}, \citenamefont {Vishwanath},\ and\ \citenamefont
  {Zaletel}}]{Bultinck:2020}%
  \BibitemOpen
  \bibfield  {author} {\bibinfo {author} {\bibfnamefont {N.}~\bibnamefont
  {Bultinck}}, \bibinfo {author} {\bibfnamefont {E.}~\bibnamefont {Khalaf}},
  \bibinfo {author} {\bibfnamefont {S.}~\bibnamefont {Liu}}, \bibinfo {author}
  {\bibfnamefont {S.}~\bibnamefont {Chatterjee}}, \bibinfo {author}
  {\bibfnamefont {A.}~\bibnamefont {Vishwanath}},\ and\ \bibinfo {author}
  {\bibfnamefont {M.~P.}\ \bibnamefont {Zaletel}},\ }\bibfield  {title}
  {\bibinfo {title} {Ground State and Hidden Symmetry of Magic-Angle Graphene
  at Even Integer Filling},\ }\href
  {https://doi.org/10.1103/PhysRevX.10.031034} {\bibfield  {journal} {\bibinfo
  {journal} {Phys. Rev. X}\ }\textbf {\bibinfo {volume} {10}},\ \bibinfo
  {pages} {031034} (\bibinfo {year} {2020})}\BibitemShut {NoStop}%
\bibitem [{\citenamefont {Kwan}\ \emph {et~al.}(2021)\citenamefont {Kwan},
  \citenamefont {Wagner}, \citenamefont {Soejima}, \citenamefont {Zaletel},
  \citenamefont {Simon}, \citenamefont {Parameswaran},\ and\ \citenamefont
  {Bultinck}}]{kwan21}%
  \BibitemOpen
  \bibfield  {author} {\bibinfo {author} {\bibfnamefont {Y.~H.}\ \bibnamefont
  {Kwan}}, \bibinfo {author} {\bibfnamefont {G.}~\bibnamefont {Wagner}},
  \bibinfo {author} {\bibfnamefont {T.}~\bibnamefont {Soejima}}, \bibinfo
  {author} {\bibfnamefont {M.~P.}\ \bibnamefont {Zaletel}}, \bibinfo {author}
  {\bibfnamefont {S.~H.}\ \bibnamefont {Simon}}, \bibinfo {author}
  {\bibfnamefont {S.~A.}\ \bibnamefont {Parameswaran}},\ and\ \bibinfo {author}
  {\bibfnamefont {N.}~\bibnamefont {Bultinck}},\ }\bibfield  {title} {\bibinfo
  {title} {Kekul\'e Spiral Order at All Nonzero Integer Fillings in Twisted
  Bilayer Graphene},\ }\href {https://doi.org/10.1103/PhysRevX.11.041063}
  {\bibfield  {journal} {\bibinfo  {journal} {Phys. Rev. X}\ }\textbf {\bibinfo
  {volume} {11}},\ \bibinfo {pages} {041063} (\bibinfo {year}
  {2021})}\BibitemShut {NoStop}%
\bibitem [{\citenamefont {Hofmann}\ \emph {et~al.}(2022)\citenamefont
  {Hofmann}, \citenamefont {Khalaf}, \citenamefont {Vishwanath}, \citenamefont
  {Berg},\ and\ \citenamefont {Lee}}]{hofmann22}%
  \BibitemOpen
  \bibfield  {author} {\bibinfo {author} {\bibfnamefont {J.~S.}\ \bibnamefont
  {Hofmann}}, \bibinfo {author} {\bibfnamefont {E.}~\bibnamefont {Khalaf}},
  \bibinfo {author} {\bibfnamefont {A.}~\bibnamefont {Vishwanath}}, \bibinfo
  {author} {\bibfnamefont {E.}~\bibnamefont {Berg}},\ and\ \bibinfo {author}
  {\bibfnamefont {J.~Y.}\ \bibnamefont {Lee}},\ }\bibfield  {title} {\bibinfo
  {title} {Fermionic Monte Carlo Study of a Realistic Model of Twisted Bilayer
  Graphene},\ }\href {https://doi.org/10.1103/PhysRevX.12.011061} {\bibfield
  {journal} {\bibinfo  {journal} {Phys. Rev. X}\ }\textbf {\bibinfo {volume}
  {12}},\ \bibinfo {pages} {011061} (\bibinfo {year} {2022})}\BibitemShut
  {NoStop}%
\bibitem [{\citenamefont {Rai}\ \emph {et~al.}(2024)\citenamefont {Rai},
  \citenamefont {Crippa}, \citenamefont {C\ifmmode \u{a}\else
  \u{a}\fi{}lug\ifmmode~\u{a}\else \u{a}\fi{}ru}, \citenamefont {Hu},
  \citenamefont {Paoletti}, \citenamefont {de' Medici}, \citenamefont
  {Georges}, \citenamefont {Bernevig}, \citenamefont {Valent\'{\i}},
  \citenamefont {Sangiovanni},\ and\ \citenamefont {Wehling}}]{rai24}%
  \BibitemOpen
  \bibfield  {author} {\bibinfo {author} {\bibfnamefont {G.}~\bibnamefont
  {Rai}}, \bibinfo {author} {\bibfnamefont {L.}~\bibnamefont {Crippa}},
  \bibinfo {author} {\bibfnamefont {D.}~\bibnamefont {C\ifmmode \u{a}\else
  \u{a}\fi{}lug\ifmmode~\u{a}\else \u{a}\fi{}ru}}, \bibinfo {author}
  {\bibfnamefont {H.}~\bibnamefont {Hu}}, \bibinfo {author} {\bibfnamefont
  {F.}~\bibnamefont {Paoletti}}, \bibinfo {author} {\bibfnamefont
  {L.}~\bibnamefont {de' Medici}}, \bibinfo {author} {\bibfnamefont
  {A.}~\bibnamefont {Georges}}, \bibinfo {author} {\bibfnamefont {B.~A.}\
  \bibnamefont {Bernevig}}, \bibinfo {author} {\bibfnamefont {R.}~\bibnamefont
  {Valent\'{\i}}}, \bibinfo {author} {\bibfnamefont {G.}~\bibnamefont
  {Sangiovanni}},\ and\ \bibinfo {author} {\bibfnamefont {T.}~\bibnamefont
  {Wehling}},\ }\bibfield  {title} {\bibinfo {title} {Dynamical Correlations
  and Order in Magic-Angle Twisted Bilayer Graphene},\ }\href
  {https://doi.org/10.1103/PhysRevX.14.031045} {\bibfield  {journal} {\bibinfo
  {journal} {Phys. Rev. X}\ }\textbf {\bibinfo {volume} {14}},\ \bibinfo
  {pages} {031045} (\bibinfo {year} {2024})}\BibitemShut {NoStop}%
\bibitem [{\citenamefont {Biedermann}\ and\ \citenamefont
  {Janssen}(2024)}]{Biedermann:2024}%
  \BibitemOpen
  \bibfield  {author} {\bibinfo {author} {\bibfnamefont {J.}~\bibnamefont
  {Biedermann}}\ and\ \bibinfo {author} {\bibfnamefont {L.}~\bibnamefont
  {Janssen}},\ }\bibinfo {title} {Twist-tuned quantum criticality in moir\'e
  bilayer graphene},\ \Eprint {https://arxiv.org/abs/2412.16042}
  {arXiv:2412.16042}\BibitemShut {NoStop}%
\bibitem [{\citenamefont {Huang}\ \emph {et~al.}(2024)\citenamefont {Huang},
  \citenamefont {Parthenios}, \citenamefont {Ulybyshev}, \citenamefont {Zhang},
  \citenamefont {Assaad}, \citenamefont {Classen},\ and\ \citenamefont
  {Meng}}]{Huang:2024}%
  \BibitemOpen
  \bibfield  {author} {\bibinfo {author} {\bibfnamefont {C.}~\bibnamefont
  {Huang}}, \bibinfo {author} {\bibfnamefont {N.}~\bibnamefont {Parthenios}},
  \bibinfo {author} {\bibfnamefont {M.}~\bibnamefont {Ulybyshev}}, \bibinfo
  {author} {\bibfnamefont {X.}~\bibnamefont {Zhang}}, \bibinfo {author}
  {\bibfnamefont {F.~F.}\ \bibnamefont {Assaad}}, \bibinfo {author}
  {\bibfnamefont {L.}~\bibnamefont {Classen}},\ and\ \bibinfo {author}
  {\bibfnamefont {Z.~Y.}\ \bibnamefont {Meng}},\ }\bibinfo {title} {Angle-Tuned
  Gross-Neveu Quantum Criticality in Twisted Bilayer Graphene: A Quantum Monte
  Carlo Study},\ \Eprint {https://arxiv.org/abs/2412.11382}
  {arXiv:2412.11382}\BibitemShut {NoStop}%
\bibitem [{\citenamefont {Vojta}(2018)}]{vojta18}%
  \BibitemOpen
  \bibfield  {author} {\bibinfo {author} {\bibfnamefont {M.}~\bibnamefont
  {Vojta}},\ }\bibfield  {title} {\bibinfo {title} {Frustration and quantum
  criticality},\ }\href {https://doi.org/10.1088/1361-6633/aab6be} {\bibfield
  {journal} {\bibinfo  {journal} {Rep. Prog. Phys.}\ }\textbf {\bibinfo
  {volume} {81}},\ \bibinfo {pages} {064501} (\bibinfo {year}
  {2018})}\BibitemShut {NoStop}%
\bibitem [{\citenamefont {Herbut}(2007)}]{Herbut:2007}%
  \BibitemOpen
  \bibfield  {author} {\bibinfo {author} {\bibfnamefont {I.}~\bibnamefont
  {Herbut}},\ }\href@noop {} {\emph {\bibinfo {title} {A modern approach to
  critical phenomena}}}\ (\bibinfo  {publisher} {Cambridge University Press},\
  \bibinfo {year} {2007})\BibitemShut {NoStop}%
\bibitem [{\citenamefont {Gracey}(2021)}]{GraceyN3XY:2021}%
  \BibitemOpen
  \bibfield  {author} {\bibinfo {author} {\bibfnamefont {J.~A.}\ \bibnamefont
  {Gracey}},\ }\bibfield  {title} {\bibinfo {title} {Critical exponent
  $\ensuremath{\eta}$ at $O(1/{N}^{3})$ in the chiral XY model using the large
  $N$ conformal bootstrap},\ }\href
  {https://doi.org/10.1103/PhysRevD.103.065018} {\bibfield  {journal} {\bibinfo
   {journal} {Phys. Rev. D}\ }\textbf {\bibinfo {volume} {103}},\ \bibinfo
  {pages} {065018} (\bibinfo {year} {2021})}\BibitemShut {NoStop}%
\bibitem [{\citenamefont {Classen}\ \emph {et~al.}(2017)\citenamefont
  {Classen}, \citenamefont {Herbut},\ and\ \citenamefont
  {Scherer}}]{ClassenXY:2017}%
  \BibitemOpen
  \bibfield  {author} {\bibinfo {author} {\bibfnamefont {L.}~\bibnamefont
  {Classen}}, \bibinfo {author} {\bibfnamefont {I.~F.}\ \bibnamefont
  {Herbut}},\ and\ \bibinfo {author} {\bibfnamefont {M.~M.}\ \bibnamefont
  {Scherer}},\ }\bibfield  {title} {\bibinfo {title} {Fluctuation-induced
  continuous transition and quantum criticality in Dirac semimetals},\ }\href
  {https://doi.org/10.1103/PhysRevB.96.115132} {\bibfield  {journal} {\bibinfo
  {journal} {Phys. Rev. B}\ }\textbf {\bibinfo {volume} {96}},\ \bibinfo
  {pages} {115132} (\bibinfo {year} {2017})}\BibitemShut {NoStop}%
\bibitem [{\citenamefont {Janssen}\ \emph {et~al.}(2018)\citenamefont
  {Janssen}, \citenamefont {Herbut},\ and\ \citenamefont
  {Scherer}}]{JanssenXYFRG:2018}%
  \BibitemOpen
  \bibfield  {author} {\bibinfo {author} {\bibfnamefont {L.}~\bibnamefont
  {Janssen}}, \bibinfo {author} {\bibfnamefont {I.~F.}\ \bibnamefont
  {Herbut}},\ and\ \bibinfo {author} {\bibfnamefont {M.~M.}\ \bibnamefont
  {Scherer}},\ }\bibfield  {title} {\bibinfo {title} {Compatible orders and
  fermion-induced emergent symmetry in Dirac systems},\ }\href
  {https://doi.org/10.1103/PhysRevB.97.041117} {\bibfield  {journal} {\bibinfo
  {journal} {Phys. Rev. B}\ }\textbf {\bibinfo {volume} {97}},\ \bibinfo
  {pages} {041117} (\bibinfo {year} {2018})}\BibitemShut {NoStop}%
\bibitem [{\citenamefont {Tolosa-Simeón}\ \emph {et~al.}(2025)\citenamefont
  {Tolosa-Simeón}, \citenamefont {Classen},\ and\ \citenamefont
  {Scherer}}]{Tolosa:2025}%
  \BibitemOpen
  \bibfield  {author} {\bibinfo {author} {\bibfnamefont {M.}~\bibnamefont
  {Tolosa-Simeón}}, \bibinfo {author} {\bibfnamefont {L.}~\bibnamefont
  {Classen}},\ and\ \bibinfo {author} {\bibfnamefont {M.~M.}\ \bibnamefont
  {Scherer}},\ }\bibinfo {title} {Relativistic Mott transitions, quantum
  criticality, and finite-temperature effects in tunable Dirac materials from
  functional renormalization},\ \Eprint {https://arxiv.org/abs/2503.04911}
  {arXiv:2503.04911}\BibitemShut {NoStop}%
\bibitem [{\citenamefont {Janssen}\ and\ \citenamefont
  {Herbut}(2014)}]{JanssenAFM:2014}%
  \BibitemOpen
  \bibfield  {author} {\bibinfo {author} {\bibfnamefont {L.}~\bibnamefont
  {Janssen}}\ and\ \bibinfo {author} {\bibfnamefont {I.~F.}\ \bibnamefont
  {Herbut}},\ }\bibfield  {title} {\bibinfo {title} {Antiferromagnetic critical
  point on graphene's honeycomb lattice: A functional renormalization group
  approach},\ }\href {https://doi.org/10.1103/PhysRevB.89.205403} {\bibfield
  {journal} {\bibinfo  {journal} {Phys. Rev. B}\ }\textbf {\bibinfo {volume}
  {89}},\ \bibinfo {pages} {205403} (\bibinfo {year} {2014})}\BibitemShut
  {NoStop}%
\bibitem [{\citenamefont {Ihrig}\ \emph {et~al.}(2018)\citenamefont {Ihrig},
  \citenamefont {Mihaila},\ and\ \citenamefont {Scherer}}]{IhrigIsing:2018}%
  \BibitemOpen
  \bibfield  {author} {\bibinfo {author} {\bibfnamefont {B.}~\bibnamefont
  {Ihrig}}, \bibinfo {author} {\bibfnamefont {L.~N.}\ \bibnamefont {Mihaila}},\
  and\ \bibinfo {author} {\bibfnamefont {M.~M.}\ \bibnamefont {Scherer}},\
  }\bibfield  {title} {\bibinfo {title} {Critical behavior of Dirac fermions
  from perturbative renormalization},\ }\href
  {https://doi.org/10.1103/PhysRevB.98.125109} {\bibfield  {journal} {\bibinfo
  {journal} {Phys. Rev. B}\ }\textbf {\bibinfo {volume} {98}},\ \bibinfo
  {pages} {125109} (\bibinfo {year} {2018})}\BibitemShut {NoStop}%
\bibitem [{\citenamefont {Ladovrechis}\ \emph {et~al.}(2023)\citenamefont
  {Ladovrechis}, \citenamefont {Ray}, \citenamefont {Meng},\ and\ \citenamefont
  {Janssen}}]{Ladovrechis:2023}%
  \BibitemOpen
  \bibfield  {author} {\bibinfo {author} {\bibfnamefont {K.}~\bibnamefont
  {Ladovrechis}}, \bibinfo {author} {\bibfnamefont {S.}~\bibnamefont {Ray}},
  \bibinfo {author} {\bibfnamefont {T.}~\bibnamefont {Meng}},\ and\ \bibinfo
  {author} {\bibfnamefont {L.}~\bibnamefont {Janssen}},\ }\bibfield  {title}
  {\bibinfo {title} {Gross-Neveu-Heisenberg criticality from
  $2+\ensuremath{\epsilon}$ expansion},\ }\href
  {https://doi.org/10.1103/PhysRevB.107.035151} {\bibfield  {journal} {\bibinfo
   {journal} {Phys. Rev. B}\ }\textbf {\bibinfo {volume} {107}},\ \bibinfo
  {pages} {035151} (\bibinfo {year} {2023})}\BibitemShut {NoStop}%
\bibitem [{\citenamefont {Gracey}\ \emph {et~al.}(2016)\citenamefont {Gracey},
  \citenamefont {Luthe},\ and\ \citenamefont {Schr\"oder}}]{gracey16}%
  \BibitemOpen
  \bibfield  {author} {\bibinfo {author} {\bibfnamefont {J.~A.}\ \bibnamefont
  {Gracey}}, \bibinfo {author} {\bibfnamefont {T.}~\bibnamefont {Luthe}},\ and\
  \bibinfo {author} {\bibfnamefont {Y.}~\bibnamefont {Schr\"oder}},\ }\bibfield
   {title} {\bibinfo {title} {Four loop renormalization of the Gross-Neveu
  model},\ }\href {https://doi.org/10.1103/PhysRevD.94.125028} {\bibfield
  {journal} {\bibinfo  {journal} {Phys. Rev. D}\ }\textbf {\bibinfo {volume}
  {94}},\ \bibinfo {pages} {125028} (\bibinfo {year} {2016})}\BibitemShut
  {NoStop}%
\bibitem [{\citenamefont {Ma}\ \emph {et~al.}(2024)\citenamefont {Ma},
  \citenamefont {Chaturvedi}, \citenamefont {Nguyen}, \citenamefont {Watanabe},
  \citenamefont {Taniguchi}, \citenamefont {Mak},\ and\ \citenamefont
  {Shan}}]{ma24}%
  \BibitemOpen
  \bibfield  {author} {\bibinfo {author} {\bibfnamefont {L.}~\bibnamefont
  {Ma}}, \bibinfo {author} {\bibfnamefont {R.}~\bibnamefont {Chaturvedi}},
  \bibinfo {author} {\bibfnamefont {P.~X.}\ \bibnamefont {Nguyen}}, \bibinfo
  {author} {\bibfnamefont {K.}~\bibnamefont {Watanabe}}, \bibinfo {author}
  {\bibfnamefont {T.}~\bibnamefont {Taniguchi}}, \bibinfo {author}
  {\bibfnamefont {K.~F.}\ \bibnamefont {Mak}},\ and\ \bibinfo {author}
  {\bibfnamefont {J.}~\bibnamefont {Shan}},\ }\bibinfo {title} {Relativistic
  Mott transition in strongly correlated artificial graphene},\ \Eprint
  {https://arxiv.org/abs/2412.07150} {arXiv:2412.07150}\BibitemShut {NoStop}%
\bibitem [{\citenamefont {Zinn-Justin}(1991)}]{zinnjustin91}%
  \BibitemOpen
  \bibfield  {author} {\bibinfo {author} {\bibfnamefont {J.}~\bibnamefont
  {Zinn-Justin}},\ }\bibfield  {title} {\bibinfo {title} {Four-fermion
  interaction near four dimensions},\ }\href
  {https://doi.org/https://doi.org/10.1016/0550-3213(91)90043-W} {\bibfield
  {journal} {\bibinfo  {journal} {Nucl. Phys. B}\ }\textbf {\bibinfo {volume}
  {367}},\ \bibinfo {pages} {105} (\bibinfo {year} {1991})}\BibitemShut
  {NoStop}%
\bibitem [{\citenamefont {Juri\ifmmode \check{c}\else
  \v{c}\fi{}i\ifmmode~\acute{c}\else \'{c}\fi{}}\ \emph
  {et~al.}(2009)\citenamefont {Juri\ifmmode \check{c}\else
  \v{c}\fi{}i\ifmmode~\acute{c}\else \'{c}\fi{}}, \citenamefont {Herbut},\ and\
  \citenamefont {Semenoff}}]{juricic09}%
  \BibitemOpen
  \bibfield  {author} {\bibinfo {author} {\bibfnamefont {V.}~\bibnamefont
  {Juri\ifmmode \check{c}\else \v{c}\fi{}i\ifmmode~\acute{c}\else \'{c}\fi{}}},
  \bibinfo {author} {\bibfnamefont {I.~F.}\ \bibnamefont {Herbut}},\ and\
  \bibinfo {author} {\bibfnamefont {G.~W.}\ \bibnamefont {Semenoff}},\
  }\bibfield  {title} {\bibinfo {title} {Coulomb interaction at the
  metal-insulator critical point in graphene},\ }\href
  {https://doi.org/10.1103/PhysRevB.80.081405} {\bibfield  {journal} {\bibinfo
  {journal} {Phys. Rev. B}\ }\textbf {\bibinfo {volume} {80}},\ \bibinfo
  {pages} {081405} (\bibinfo {year} {2009})}\BibitemShut {NoStop}%
\bibitem [{\citenamefont {Gies}\ and\ \citenamefont
  {Janssen}(2010)}]{JanssenThirring:2010}%
  \BibitemOpen
  \bibfield  {author} {\bibinfo {author} {\bibfnamefont {H.}~\bibnamefont
  {Gies}}\ and\ \bibinfo {author} {\bibfnamefont {L.}~\bibnamefont {Janssen}},\
  }\bibfield  {title} {\bibinfo {title} {UV fixed-point structure of the
  three-dimensional Thirring model},\ }\href
  {https://doi.org/10.1103/PhysRevD.82.085018} {\bibfield  {journal} {\bibinfo
  {journal} {Phys. Rev. D}\ }\textbf {\bibinfo {volume} {82}},\ \bibinfo
  {pages} {085018} (\bibinfo {year} {2010})}\BibitemShut {NoStop}%
\bibitem [{\citenamefont {Braun}\ \emph {et~al.}(2011)\citenamefont {Braun},
  \citenamefont {Gies},\ and\ \citenamefont {Scherer}}]{braun11}%
  \BibitemOpen
  \bibfield  {author} {\bibinfo {author} {\bibfnamefont {J.}~\bibnamefont
  {Braun}}, \bibinfo {author} {\bibfnamefont {H.}~\bibnamefont {Gies}},\ and\
  \bibinfo {author} {\bibfnamefont {D.~D.}\ \bibnamefont {Scherer}},\
  }\bibfield  {title} {\bibinfo {title} {Asymptotic safety: A simple example},\
  }\href {https://doi.org/10.1103/PhysRevD.83.085012} {\bibfield  {journal}
  {\bibinfo  {journal} {Phys. Rev. D}\ }\textbf {\bibinfo {volume} {83}},\
  \bibinfo {pages} {085012} (\bibinfo {year} {2011})}\BibitemShut {NoStop}%
\bibitem [{\citenamefont {Boyack}\ \emph {et~al.}(2019)\citenamefont {Boyack},
  \citenamefont {Rayyan},\ and\ \citenamefont {Maciejko}}]{boyack19}%
  \BibitemOpen
  \bibfield  {author} {\bibinfo {author} {\bibfnamefont {R.}~\bibnamefont
  {Boyack}}, \bibinfo {author} {\bibfnamefont {A.}~\bibnamefont {Rayyan}},\
  and\ \bibinfo {author} {\bibfnamefont {J.}~\bibnamefont {Maciejko}},\
  }\bibfield  {title} {\bibinfo {title} {Deconfined criticality in the
  ${\mathrm{QED}}_{3}$ Gross-Neveu-Yukawa model: The $1/N$ expansion
  revisited},\ }\href {https://doi.org/10.1103/PhysRevB.99.195135} {\bibfield
  {journal} {\bibinfo  {journal} {Phys. Rev. B}\ }\textbf {\bibinfo {volume}
  {99}},\ \bibinfo {pages} {195135} (\bibinfo {year} {2019})}\BibitemShut
  {NoStop}%
\bibitem [{\citenamefont {Erramilli}\ \emph {et~al.}(2023)\citenamefont
  {Erramilli}, \citenamefont {Iliesiu}, \citenamefont {Kravchuk}, \citenamefont
  {Liu}, \citenamefont {Poland},\ and\ \citenamefont
  {Simmons-Duffin}}]{Erramilli:2022kgp}%
  \BibitemOpen
  \bibfield  {author} {\bibinfo {author} {\bibfnamefont {R.~S.}\ \bibnamefont
  {Erramilli}}, \bibinfo {author} {\bibfnamefont {L.~V.}\ \bibnamefont
  {Iliesiu}}, \bibinfo {author} {\bibfnamefont {P.}~\bibnamefont {Kravchuk}},
  \bibinfo {author} {\bibfnamefont {A.}~\bibnamefont {Liu}}, \bibinfo {author}
  {\bibfnamefont {D.}~\bibnamefont {Poland}},\ and\ \bibinfo {author}
  {\bibfnamefont {D.}~\bibnamefont {Simmons-Duffin}},\ }\bibfield  {title}
  {\bibinfo {title} {The Gross-Neveu-Yukawa archipelago},\ }\href
  {https://doi.org/10.1007/JHEP02(2023)036} {\bibfield  {journal} {\bibinfo
  {journal} {J. High Energy Phys.}\ }2\bibfield  {volume} {\bibinfo  {volume} {
  (2023)}\ }36}\BibitemShut {NoStop}%
\bibitem [{\citenamefont {Janssen}(2016)}]{JanssenLorentz:2016}%
  \BibitemOpen
  \bibfield  {author} {\bibinfo {author} {\bibfnamefont {L.}~\bibnamefont
  {Janssen}},\ }\bibfield  {title} {\bibinfo {title} {Spontaneous breaking of
  Lorentz symmetry in ($2+\ensuremath{\epsilon}$)-dimensional QED},\ }\href
  {https://doi.org/10.1103/PhysRevD.94.094013} {\bibfield  {journal} {\bibinfo
  {journal} {Phys. Rev. D}\ }\textbf {\bibinfo {volume} {94}},\ \bibinfo
  {pages} {094013} (\bibinfo {year} {2016})}\BibitemShut {NoStop}%
\bibitem [{\citenamefont {Di~Pietro}\ \emph {et~al.}(2016)\citenamefont
  {Di~Pietro}, \citenamefont {Komargodski}, \citenamefont {Shamir},\ and\
  \citenamefont {Stamou}}]{dipietro16}%
  \BibitemOpen
  \bibfield  {author} {\bibinfo {author} {\bibfnamefont {L.}~\bibnamefont
  {Di~Pietro}}, \bibinfo {author} {\bibfnamefont {Z.}~\bibnamefont
  {Komargodski}}, \bibinfo {author} {\bibfnamefont {I.}~\bibnamefont
  {Shamir}},\ and\ \bibinfo {author} {\bibfnamefont {E.}~\bibnamefont
  {Stamou}},\ }\bibfield  {title} {\bibinfo {title} {Quantum Electrodynamics in
  $d=3$ from the $\ensuremath{\epsilon}$ Expansion},\ }\href
  {https://doi.org/10.1103/PhysRevLett.116.131601} {\bibfield  {journal}
  {\bibinfo  {journal} {Phys. Rev. Lett.}\ }\textbf {\bibinfo {volume} {116}},\
  \bibinfo {pages} {131601} (\bibinfo {year} {2016})}\BibitemShut {NoStop}%
\bibitem [{\citenamefont {Benvenuti}\ and\ \citenamefont
  {Khachatryan}(2019)}]{benvenuti19}%
  \BibitemOpen
  \bibfield  {author} {\bibinfo {author} {\bibfnamefont {S.}~\bibnamefont
  {Benvenuti}}\ and\ \bibinfo {author} {\bibfnamefont {H.}~\bibnamefont
  {Khachatryan}},\ }\bibfield  {title} {\bibinfo {title} {Easy-plane QED$_3$'s
  in the large $N_f$ limit},\ }\href {https://doi.org/10.1007/JHEP05(2019)214}
  {\bibfield  {journal} {\bibinfo  {journal} {J. High Energy Phys.}\
  }5\bibfield  {volume} {\bibinfo  {volume} { (2019)}\ }214}\BibitemShut
  {NoStop}%
\bibitem [{\citenamefont {Gies}\ and\ \citenamefont
  {Wetterich}(2002)}]{gies02}%
  \BibitemOpen
  \bibfield  {author} {\bibinfo {author} {\bibfnamefont {H.}~\bibnamefont
  {Gies}}\ and\ \bibinfo {author} {\bibfnamefont {C.}~\bibnamefont
  {Wetterich}},\ }\bibfield  {title} {\bibinfo {title} {Renormalization flow of
  bound states},\ }\href {https://doi.org/10.1103/PhysRevD.65.065001}
  {\bibfield  {journal} {\bibinfo  {journal} {Phys. Rev. D}\ }\textbf {\bibinfo
  {volume} {65}},\ \bibinfo {pages} {065001} (\bibinfo {year}
  {2002})}\BibitemShut {NoStop}%
\bibitem [{\citenamefont {Pawlowski}(2007)}]{Pawlowski:2005xe}%
  \BibitemOpen
  \bibfield  {author} {\bibinfo {author} {\bibfnamefont {J.~M.}\ \bibnamefont
  {Pawlowski}},\ }\bibfield  {title} {\bibinfo {title} {{Aspects of the
  functional renormalisation group}},\ }\href
  {https://doi.org/10.1016/j.aop.2007.01.007} {\bibfield  {journal} {\bibinfo
  {journal} {Annals Phys.}\ }\textbf {\bibinfo {volume} {322}},\ \bibinfo
  {pages} {2831} (\bibinfo {year} {2007})}\BibitemShut {NoStop}%
\bibitem [{\citenamefont {Janssen}\ and\ \citenamefont
  {Herbut}(2017)}]{janssen17}%
  \BibitemOpen
  \bibfield  {author} {\bibinfo {author} {\bibfnamefont {L.}~\bibnamefont
  {Janssen}}\ and\ \bibinfo {author} {\bibfnamefont {I.~F.}\ \bibnamefont
  {Herbut}},\ }\bibfield  {title} {\bibinfo {title} {Phase diagram of
  electronic systems with quadratic Fermi nodes in $2<d<4$: $2 + \epsilon$
  expansion, $4 - \epsilon$ expansion, and functional renormalization group},\
  }\href {https://doi.org/10.1103/PhysRevB.95.075101} {\bibfield  {journal}
  {\bibinfo  {journal} {Phys. Rev. B}\ }\textbf {\bibinfo {volume} {95}},\
  \bibinfo {pages} {075101} (\bibinfo {year} {2017})}\BibitemShut {NoStop}%
\bibitem [{\citenamefont {Moser}\ and\ \citenamefont
  {Janssen}(2024)}]{moser24}%
  \BibitemOpen
  \bibfield  {author} {\bibinfo {author} {\bibfnamefont {D.~J.}\ \bibnamefont
  {Moser}}\ and\ \bibinfo {author} {\bibfnamefont {L.}~\bibnamefont
  {Janssen}},\ }\bibinfo {title} {Continuous order-to-order quantum phase
  transitions from fixed-point annihilation},\ \Eprint
  {https://arxiv.org/abs/2412.06890} {arXiv:2412.06890}\BibitemShut {NoStop}%
\end{thebibliography}%

\end{document}